\documentclass[aps,10pt,twocolumn,floatfix,superscriptaddress,showpacs,preprintnumbers,amsmath,amssymb,showpacs,showkeys]{revtex4}

\usepackage[utf8]{inputenc}

\usepackage{amssymb}
\usepackage{amsmath}
\usepackage{amsfonts}
\usepackage{tabularx}
\usepackage{cancel} 
\usepackage{units} 

\usepackage{epsfig} 
\usepackage{graphicx}
\usepackage{subfigure}
\usepackage{placeins}
\usepackage{grffile}
\usepackage{booktabs}

\usepackage{enumitem}

\newcommand{\seq}{\begin{subequations}}
\newcommand{\sen}{\end{subequations}}
\newcommand{\eq}{\begin{eqnarray}}
\newcommand{\en}{\end{eqnarray}}

\def\shiftdown#1{#1\llap{\lower.04ex\hbox{#1}}}

\newcommand{\ra}{\rangle}

\def\L2{\Lambda^2}

\def\jp{{J/\psi}}

\def\eq{\begin{eqnarray}}
\def\en{\end{eqnarray}}
\def\zc{Z_c^+}

\def\zcp{Z_c^{'+}}
\def\zb{Z_b^+}
\def\zbp{Z_b^{'+}}
\def\gam{\gamma}
\def\ep{e^+}
\def\em{e^-}
\def\mup{\mu^+}
\def\mum{\mu^-}
\def\transGam{\zc\to\jp\pi^+\gam}
\def\transepem{\zc\to\jp\pi^+\ep\em}
\def\transmupmum{\zc\to\jp\pi^+\mup\mum}
\def\translplm{\zc\to\jp\pi^+\ell^+\ell^-}
\def\transGamp{\zcp\to\jp\pi^+\gam}

\def\translplmp{\zcp\to\jp\pi^+\ell^+\ell^-}

\begin{document}
\unitlength = 1mm

\title{Radiative and dilepton decays of the 
hadronic molecule $Z_c^+(3900)$}

\author{Thomas Gutsche}
\affiliation{Institut f\"ur Theoretische Physik,
Universit\"at T\"ubingen, \\
Kepler Center for Astro and Particle Physics, \\ 
Auf der Morgenstelle 14, D-72076 T\"ubingen, Germany}
\author{Matthias Kesenheimer}
\affiliation{Institut f\"ur Theoretische Physik,
Universit\"at T\"ubingen, \\
Kepler Center for Astro and Particle Physics, \\ 
Auf der Morgenstelle 14, D-72076 T\"ubingen, Germany}
\author{Valery E. Lyubovitskij} 
\affiliation{Institut f\"ur Theoretische Physik,
Universit\"at T\"ubingen, \\
Kepler Center for Astro and Particle Physics, \\
Auf der Morgenstelle 14, D-72076 T\"ubingen, Germany}
\affiliation{Department of Physics, Tomsk State University,  
634050 Tomsk, Russia} 
\affiliation{Mathematical Physics Department,
Tomsk Polytechnic University, \\
Lenin ave. 30, 634050 Tomsk, Russia}

\hspace*{.2cm}
\date{\today}

\begin{abstract}

The newly observed hidden-charm meson $\zc(3900)$ with quantum numbers
$J^{P} = 1^{+}$ is considered as a hadronic molecule composed of 
$\bar{D}D^\ast$. We give detailed predictions for the decay modes 
$\transGam$ and $\translplm$ $(\ell = e, \mu)$ 
in a phenomenological Lagrangian approach.

\end{abstract}



\maketitle

\section{Introduction}
\label{intro}

Recently the three experimental collaborations
BESIII~\cite{Ablikim:2013mio}, Belle~\cite{Liu:2013dau} and
CLEO-c~\cite{Xiao:2013iha} reported on the observation of 
a new charged resonance $Z_c(3900)$ detected via the decay channel 
$J/\psi \pi^{\pm}$.
This observation of a charged, hidden charm state embedded in the
charmonium spectrum presents clear evidence for an exotic meson
resonance. Interpretations of this unusual state were immediately presented,
dominantly focusing on either a compact tetraquark
configuration~\cite{Faccini:2013lda,Dias:2013xfa,Braaten:2013boa} 
or a molecular 
state~\cite{Wang:2013cya,Cui:2013yva,Zhang:2013aoa,Ke:2013gia,Dong:2013iqa}.

One of the tools in identifying the underlying structure rests on the study
of the decay patterns of the $Z_c$ in addition to the discovery decay mode 
$J/\psi \pi^{\pm}$. For example, predictions for the strong two-body
transitions $Z_c^+ \to  H + \pi^+$ with $H = \Psi (nS)$, $h_c(mP)$ were 
already worked out in the context of a hadronic molecule 
interpretation~\cite{Dong:2013iqa} (see also extension on bottom sector 
$Z_b$~\cite{Dong:2012hc}).  
In the present work we extend to the radiative and dilepton decays of the 
hadronic molecule $Z_c^+(3900)$ proceeding as $\transGam$ and 
$\translplm$ $(\ell = e, \mu)$. 
Radiative and dilepton decays can shed light on the composite structure 
of the $Z_c^+(3900)$: decay patterns and decay widths will depend on
the structure assumption such as a hadronic molecule, a tetraquark 
configuration or even a mixed state of these components. 
Although radiative decays are suppressed due to the strength of the 
interaction particular final states such as $J/\psi \pi^+ \gamma $ 
are relatively easy to identify experimentally. 
The analysis is based on a phenomenological Lagrangian 
approach~\cite{Faessler:2007gv,Dong:2012hc,Dong:2013kta} 
together with the compositeness 
condition~\cite{Weinberg:1962hj,Efimov:1993ei,Branz:2009cd},
which is a powerful tool to formulate hadrons as bound states of their 
constituents using methods of quantum field theory.
       
When treating the $\zc$ as a hadronic molecule we assume that 
this state together with the negative $Z_c(3900)^-$ and the neutral 
$Z_c(3900)^0$ partners form an isospin triplet
with spin and parity quantum numbers $J^P = 1^{+}$ 
[as for example briefly discussed in Ref.~\cite{Dong:2013kta}]. 
Therefore the charged hidden-charm meson resonance $\zc$ is set up as a
superposition of the molecular configurations $\bar{D}D^{*}$ as
\eq
|Z_c^+\ra &=&\frac{1}{\sqrt{2}} 
\Big| D^{*+}\bar{D}^0+\bar{D}^{*0}D^+ \Big\ra \,. 
\en
Note that analogous states in the bottom sector ($\zb$ and $\zbp$)
have also been considered previously Ref.~\cite{Dong:2012hc}.

To evaluate the radiative and dilepton decays of the molecular state
$\zc$ we proceed in the present paper as follows. In Sec.~\ref{framework} 
we briefly review the basic ideas of our approach. We set up the 
new resonance $\zc$ as a $\bar{D}D^\ast$ molecular state and specify 
the relevant interaction Lagrangians for the description of the three- and
four-body decays $\transGam$ and $\translplm$ $(\ell = e, \mu)$.  
In Sec.~\ref{kinematics} we introduce and discuss the kinematics of
the many-body transitions. In Sec.~\ref{results} we present numerical 
results and the discussion.

\section{Framework}
\label{framework}

\begin{figure}[hb]
	\centering
	\includegraphics[scale=.45]{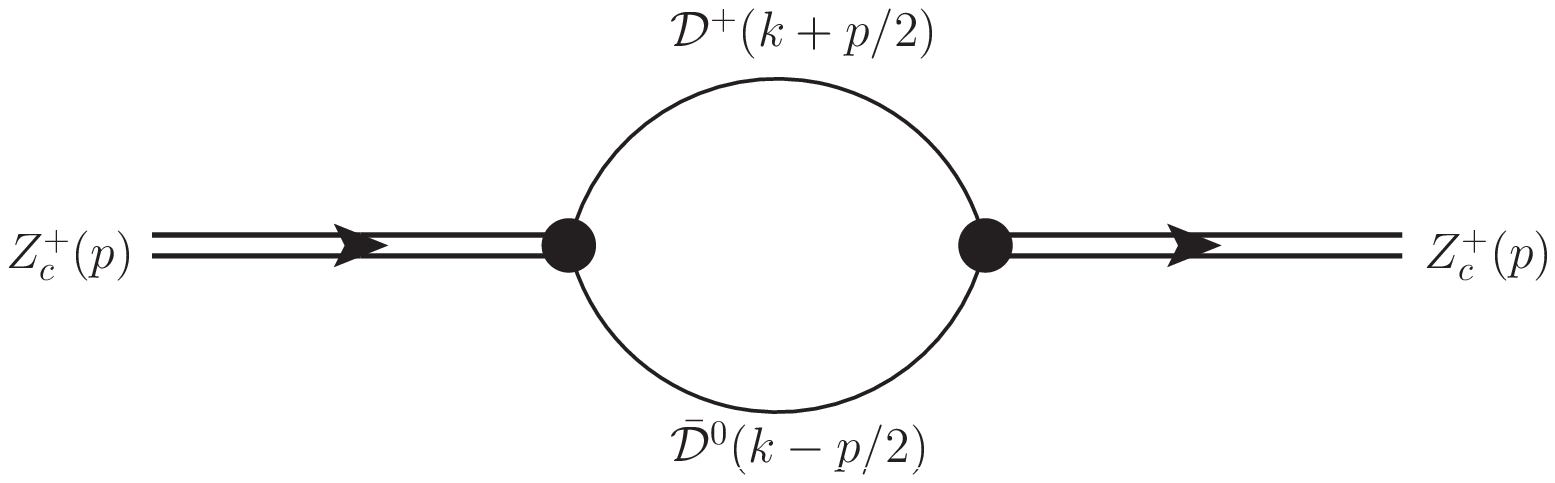}
	\caption{Mass operator for the $Z_c$. 
}
	\label{fig:massoperator}
\end{figure}

Our approach to the molecular $\zc$ state is based on an interaction
Lagrangian describing its coupling to the 
constituents as 
\eq\label{Lagr} 
\hspace*{-.6cm}
{\cal L}^{0}_{Z_c DD^\ast }(x)&=&Z_c^{+ \mu}(x) \bar J_{Z_c}^\mu(x) + 
Z_c^{- \mu}(x) J_{Z_c}^\mu(x) \,, \nonumber\\
\hspace*{-.6cm}
J_{Z_c}^\mu(x) &=& \frac{g_{_{Z_c}}}{\sqrt{2}}\, M_{Z_c} \, 
\int d^4y \, \Phi_{Z_c}(y^2)\nonumber\\ 
\hspace*{-.6cm}
&\times& \left(D^+(x_+)
\bar{D}^{*0}_{\mu}(x_-)+
D^{*+}_{\mu}(x_+)\bar{D^0}(x_-) \right) 
\en
where $x_+ \equiv x+ w_1\, y$, $x_- \equiv x- w_2\, y$. 
Here $x$ is the center-of-mass coordinate, $y$ is the relative 
(Jacobi) coordinate of the constituents and  
$w_1 = m_D/(m_D + m_{D^\ast})$ and $w_2 = m_{D^\ast}/(m_D + m_{D^\ast})$ are 
the fractions of the masses of the constituents.
The dimensionless constant $g_{_{Z_c}}$ describes the coupling strength 
of the $\zc$ to the molecular $\bar{D}D^{*}$ components. 
$\Phi_{Z_c}(y^2)$ is a correlation function 
which describes the distribution of the constituent mesons in the 
bound state. A basic requirement for the choice of an explicit form 
of the correlation function $\Phi_{Z_c}(y^2)$ 
is that its Fourier transform vanishes sufficiently fast in the ultraviolet
region of Euclidean space to render the Feynman diagrams ultraviolet finite.
We adopt a Gaussian form for the correlation function.
The Fourier transform of this vertex function is given by
$\tilde\Phi_{Z_c}(p_E^2/\Lambda^2) \equiv \exp( - p_E^2/\Lambda^2)$, 
where $p_{E}$ is the Euclidean Jacobi momentum.
$\Lambda$ is a size parameter characterizing the distribution of the two
constituent mesons in the $\zc$ system.
For a molecular system where the binding energy is negligible in
comparison with the constituent masses this size parameter is
expected to be smaller than 1 GeV. 
From our previous analyses of the strong two-body decays of the
$X, Y, Z$ meson resonances and of the $\Lambda_c(2940)$ and $\Sigma_c(2880)$
baryon states we deduced a value of maximally
$\Lambda \sim 1$~GeV~\cite{Dong:2009tg}.
For a very loosely bound system like the $X(3872)$ a size parameter of
$\Lambda \sim 0.5$~GeV~\cite {Dong:2009uf} is more suitable.
Here we choose values for $\Lambda $ in the range of $0.5-0.75$ GeV which
reflect a weakly bound heavy meson system.
Once $\Lambda$ is fixed the coupling constant $g_{_{Z_c}}$ 
is then determined by the compositeness 
condition~\cite{Dong:2013iqa,Faessler:2007gv,Branz:2009cd}. 
It implies that the renormalization constant 
of the hadron wave function is set equal to zero with:
\eq\label{ZLc}
Z_{Z_c}  = 1 - \Sigma_{Z_c}^\prime(M_{Z_c}^2) = 0 \,.
\en
$\Sigma_{Z_c}^\prime$ is the derivative of the 
transverse part of the mass operator $\Sigma_{Z_c}^{\mu\nu}$ 
of the molecular states [see Fig. \ref{fig:massoperator}], 
which is defined as
\eq
\Sigma_{Z_c}^{\mu\nu}(p) = \biggl(g^{\mu\nu} - \frac{p^\mu p^\nu}{p^2}\biggr) \, 
\Sigma_{Z_c}(p)
+ \frac{p^\mu p^\nu}{p^2}
\Sigma_{Z_c}^L(p)
\en
We would like to stress again, that the size effects of 
the constituent $D$ mesons in the $Z_c^+(3900)$ are taken 
into account by the dimensional parameter $\Lambda$ 
and the coupling constant $g_{Z_c}$. 
These parameters are constrained by the compositeness 
condition (4) --- the key condition for a study of bound states 
in quantum field theory. It is equivalent to the normalization of
the wave function in Bethe-Salpeter approaches and to the Ward identity,
relating hadronic electromagnetic vertex functions to the derivative of their 
mass operators on the mass-shell (see detailed discussion 
in Refs.~\cite{Weinberg:1962hj,Efimov:1993ei,Branz:2009cd}). 

\begin{figure}
	\centering
	\includegraphics[scale=.47]{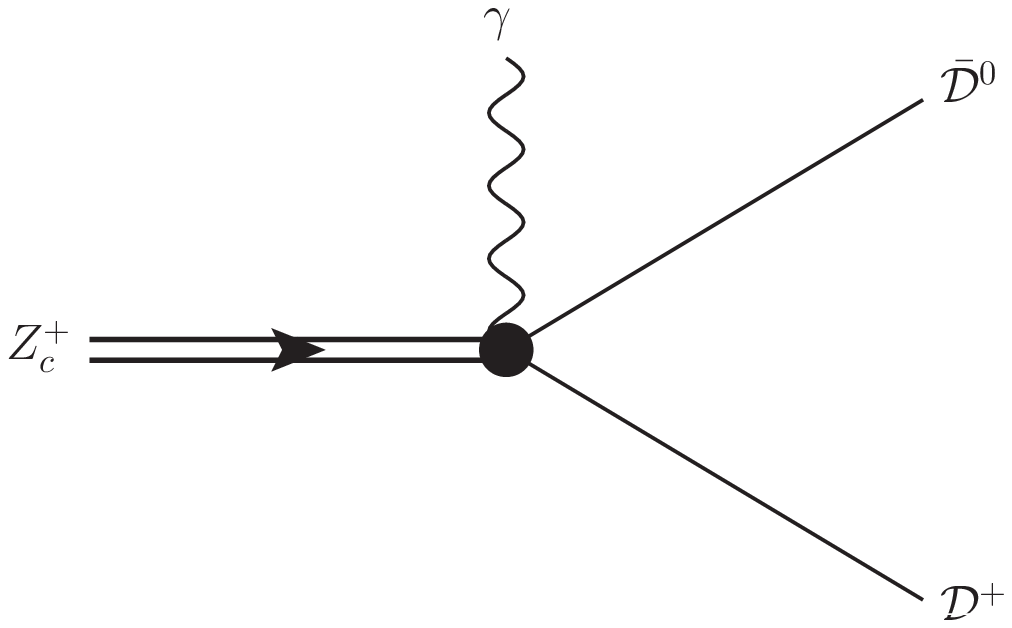}
	\caption{Photon attached to the nonlocal vertex.}
	\label{fig:feyn_nonlocalPhoton}
\centering
\includegraphics[scale=.47]{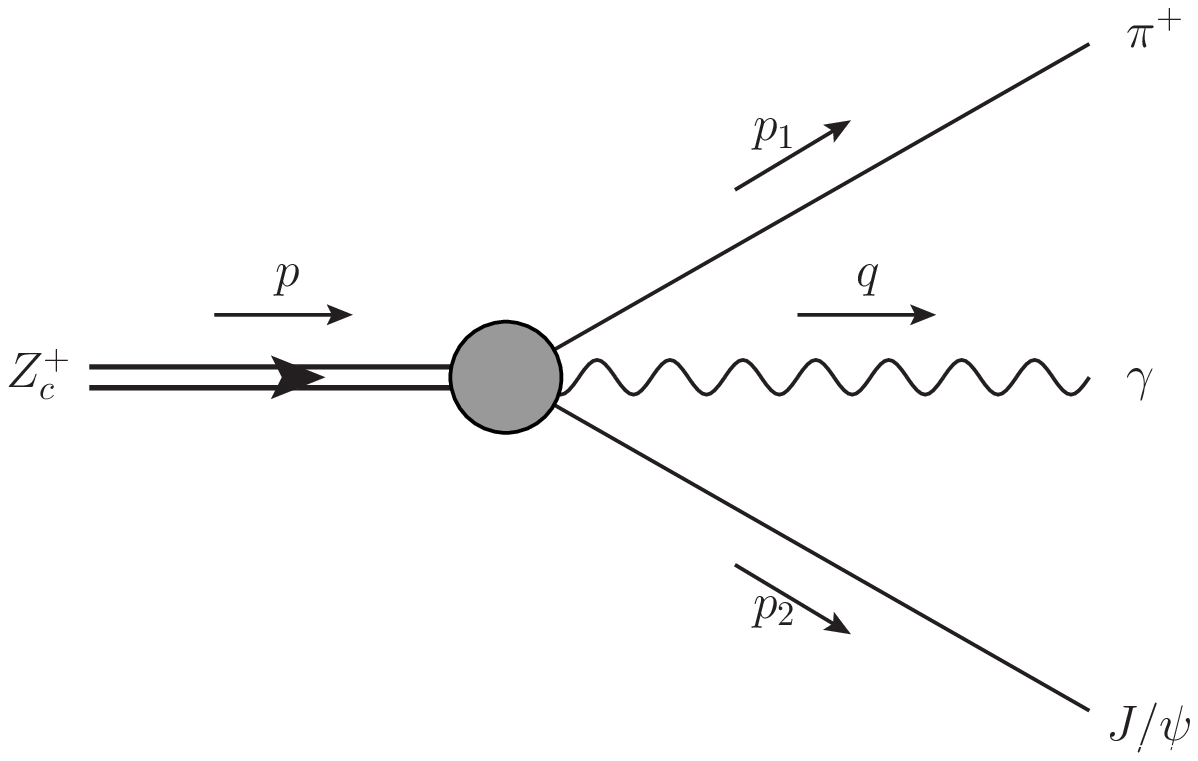}
\caption{Notation for the kinematics of the process\\ 
$\zc\to \jp\pi^+\gam$.}
\label{fig:feyn_gen1}
\centering
\includegraphics[scale=.47]{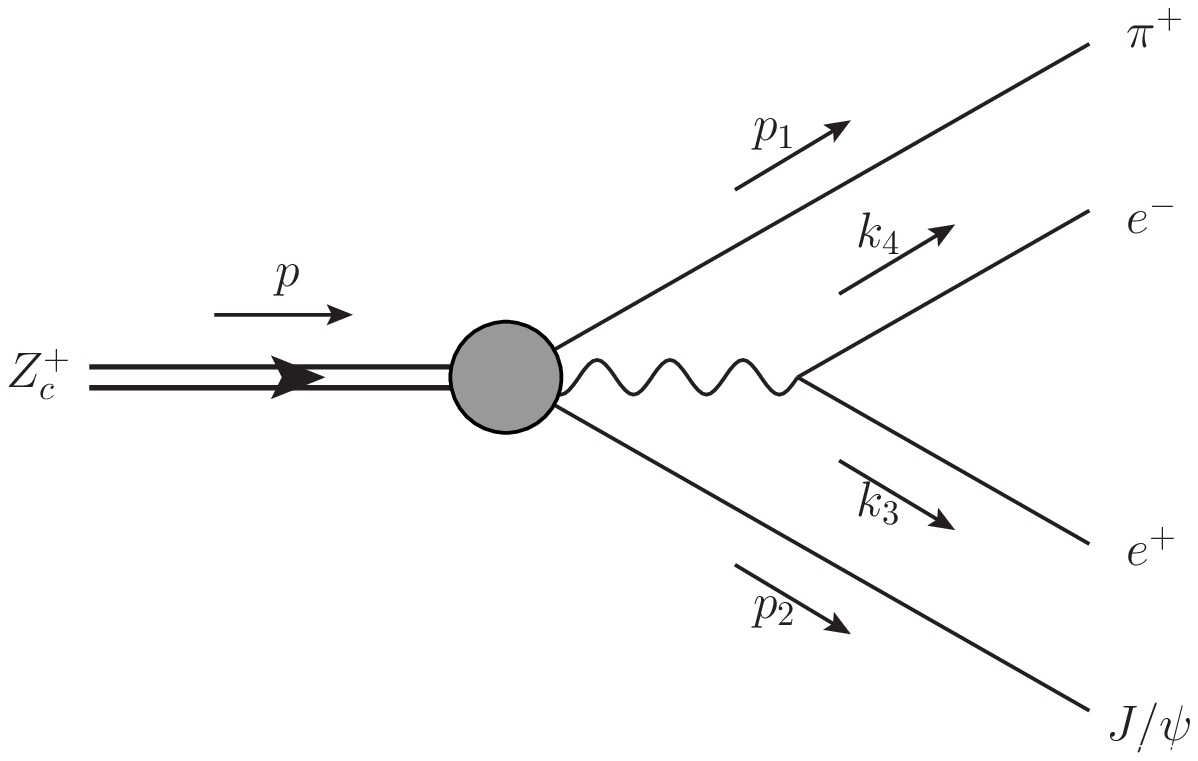}
\caption{Notation for the kinematics of the process\\ 
$\zc\to \jp\pi^+ \gamma[\to \ep \em]$.}
\label{fig:feyn_gen2}
\end{figure}

An analytical expression for the coupling $g_{_{Z_c}}$ 
is given in Appendix~\ref{massop}. In the calculation 
the mass of the $Z_c$ is expressed in terms of the constituent 
masses and the binding energy $\epsilon$ 
(a variable quantity in our calculations):
$m_{Z_c}=m_{D}+m_{D^\ast}-\epsilon$\,  
where $\epsilon$ is the binding energy.
In order to calculate the three- and four-body decays  
$\transGam$ and $\translplm$ $(\ell = e, \mu)$
we need to specify additional phenomenological Lagrangians. Their
interaction vertices occur in meson-loop diagrams which generate
the decay modes. 
For this purpose the lowest-order Lagrangian $\cal L$ is given by
\eq
{\cal L}(x) &=& {\cal L} _{Z_c}(x) + {\cal L}_{Z_cDD^\ast }(x) 
             + {\cal L}_{DD^\ast\jp\pi}(x) \nonumber \\
            &+& {\cal L}_{D}(x) + {\cal L}_{D^\ast}(x) 
             + {\cal L}_{\pi}(x) + {\cal L}_{\jp}(x)
\en
where
\eq
{\cal L}_{V}(x) &=& -\frac{1}{4} G_{\mu \nu}(x) G^{\mu \nu}(x) 
                   - \frac{1}{2} M_{V}^2 V_{\mu}(x) V^{\mu}(x)\,, 
\nonumber\\
{\cal L}_{S}(x) &=& \frac{1}{2} (D^\mu S(x))^2-\frac{1}{2} m_S^2 S^2(x)  
\en
are the free Lagrangians of the spin-1 mesons $V = Z_c, D^\ast,\jp$ 
and spin-0 mesons $S = \pi, D$, respectively; 
$G^{\mu \nu} =  \nabla^\mu V^\nu-\nabla^\nu V^\mu$ 
is the stress tensor of vector/axial mesons, 
$\nabla^\mu = \partial^\mu - ie A^\mu$ is the covariant derivative 
including the electromagnetic field in case of charged 
states and $e$ is the corresponding electric charge of hadron $H$. 
The Lagrangian ${\cal L}_{Z_cDD^\ast }(x)$ is the gauge-invariant extension 
of the strong $\zc D D^\ast$ interaction Lagrangian~(\ref{Lagr}). 
It includes photons by gauging with the path integral 
$I(x,y) = \int_{x}^{y}A_{\mu}dw^{\mu}$ resulting in 
\eq 
\label{GILagrangian}
	{\cal L}_{Z_cDD^\ast }(x) &=&\frac{g_{_{Z_c}}}{\sqrt{2}}\, M_{Z_c} \,
	Z_c^{- \mu}(x) \!\int\! d^4y \Phi_{Z_c}(y^2) 
e^{-i e I(x_+,x)}\nonumber\\
&\times&
\left(D^+(x_+) \bar{D}^{*0}_{\mu}(x_-) 
+ D^{*+}_{\mu}(x_+)\bar{D^0}(x_-) \right)\nonumber\\
&+& {\rm H.c.} 
\en 
This Lagrangian is manifestly gauge-invariant. 
As discussed in detail in Refs.~\cite{Anikin:1995cf}-\cite{Faessler:2006ky}, 
the presence of the vertex form factor in the interaction 
Lagrangian [like the strong-interaction Lagrangian describing 
the coupling of $Z_c$ to its constituents (Eq.~\ref{Lagr})] 
requires special care in establishing gauge invariance. 
One of the possibilities is provided by a modification of 
the charged fields which are multiplied by an exponential 
containing the electromagnetic field and the electrical charge $e$. 
This procedure was first suggested in Ref.~\cite{Mandelstam:1962mi} 
and applied in Refs.~\cite{Terning:1991yt,Faessler:2007gv,
Faessler:2007us,Faessler:2007cu,Faessler:2003yf}. 
In doing so the fields of the charged ${\cal D}=D,D^\ast$ mesons 
are modified as
\eq
{\cal D}^+(x_+) \ \rightarrow  \ e^{-i e I(x_+,x,P)} {\cal D}^+(x_+) \,,
\en  
which leads to the electromagnetic gauge-invariant 
Lagrangian (\ref{GILagrangian}). 
The interacting terms up to first order in $A^{\mu}$ 
are obtained by expanding ${\cal L}_{Z_cDD^\ast }(x)$ 
in terms of $I(x_+,x)$. Diagrammatically the first order 
term gives rise to a nonlocal vertex with an additional 
photon line attached [see Fig.~\ref{fig:feyn_nonlocalPhoton}]. 

In the calculation of the three- and four-body decays 
$\transGam$ and $\translplm$ $(\ell = e, \mu)$ 
we also need the four-particle $DD^\ast \jp \pi$ vertex 
generated by a phenomenological Lagrangian proposed 
in Ref.~\cite{Dong:2013iqa}
\eq 
\label{phenLagrangian}
  {\cal L}_{DD^\ast\jp\pi}(x)&=& - g_{DD^\ast\jp\pi} 
J^{\mu\nu}(x) \, \bar D^\ast_\nu(x) \,
  \nabla_\mu \hat{\pi}(x) \, D(x) \nonumber\\
&+& {\rm H.c.}
\en 
where $\nabla_\mu$ is the covariant derivative. 
The coupling constant $g_{DD^\ast\jp\pi}$ is given by
\eq
g_{DD^\ast\jp\pi}  = 
\frac{\sqrt{6}}{2\sqrt{2}}\frac{g_{J D D} g_{D^\ast D \pi}}{(m_{Z_c}^2-m_{\jp}^2) 
(1+\frac{m_{\jp}^2}{2 m_{Z_c}^2})}\, ,
\en  
as defined in Ref.~\cite{Dong:2013iqa}. 
The numerical value for $g_{J D D} g_{D^\ast D \pi}/2 \sqrt{2}$ was evaluated in 
Ref.~\cite{Dong:2013iqa} 
and is expressed through $g_{J D D} g_{D^\ast D \pi}/2 \sqrt{2} = 47.08$. 
$\hat{\pi}$ is the pion matrix 
\eq
  \hat{\pi} = \vec{\pi} \cdot \vec{\tau} =
  \begin{pmatrix}
 \pi^0         & \pi^+\sqrt{2} \\
 \pi^-\sqrt{2} & - \pi^0 \\
  \end{pmatrix}
\en
and $J^{\mu\nu} = \partial^\mu J^\nu-\partial^\nu J^\mu$ 
is the stress tensor of the $\jp$ state. $D=(D^+,D^0)$, 
$D^\ast=(D^{*+},D^{*0})$ are the doublets of pseudoscalar 
and vector charmed $D$ mesons, respectively. 
To simplify the calculations we neglect the transverse part 
of the vector propagators and of all 
vertex factors where the vector fields are involved. 
This is justified by the fact that the transverse 
parts only give a minor contribution to the transition amplitude.
For a better and compact overview the external momenta of the
three- and four-body decays are summarized in Figs. 3
and 4, respectively.
To summarize we also indicate the respective Feynman rules 
in Appendix~\ref{feynrules}.

\begin{figure}[htb]
	\centering
	\includegraphics[scale=.44]{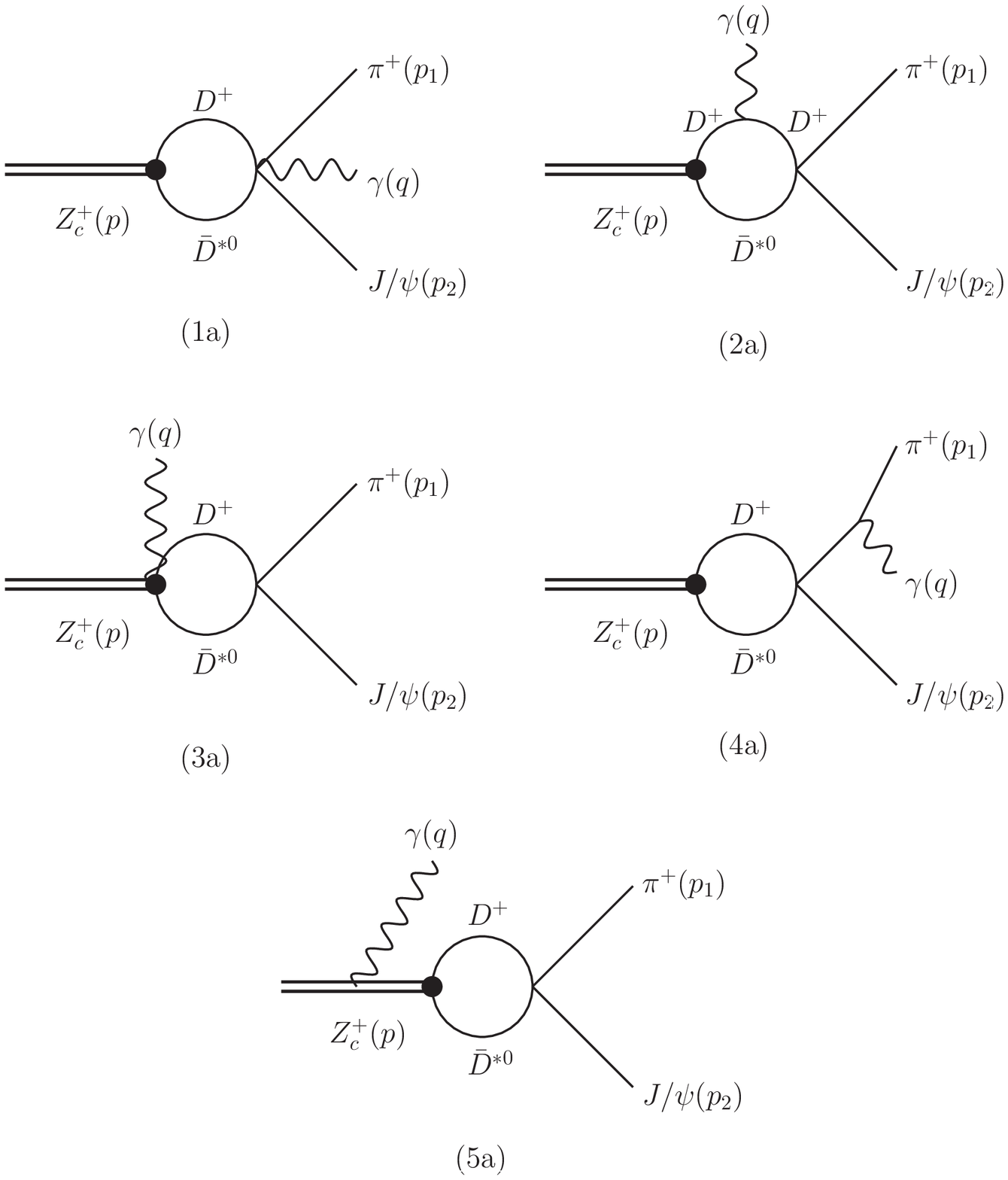}
	\caption{$\bar{D^{*0}} D^+$ meson loop diagrams contributing 
        to the three-body decay $\zc\to \jp\pi^+\gam$.}
	\label{diagramsZcJpsPiGamDsD1}
\vspace*{.25cm}
	\centering
	\includegraphics[scale=.44]{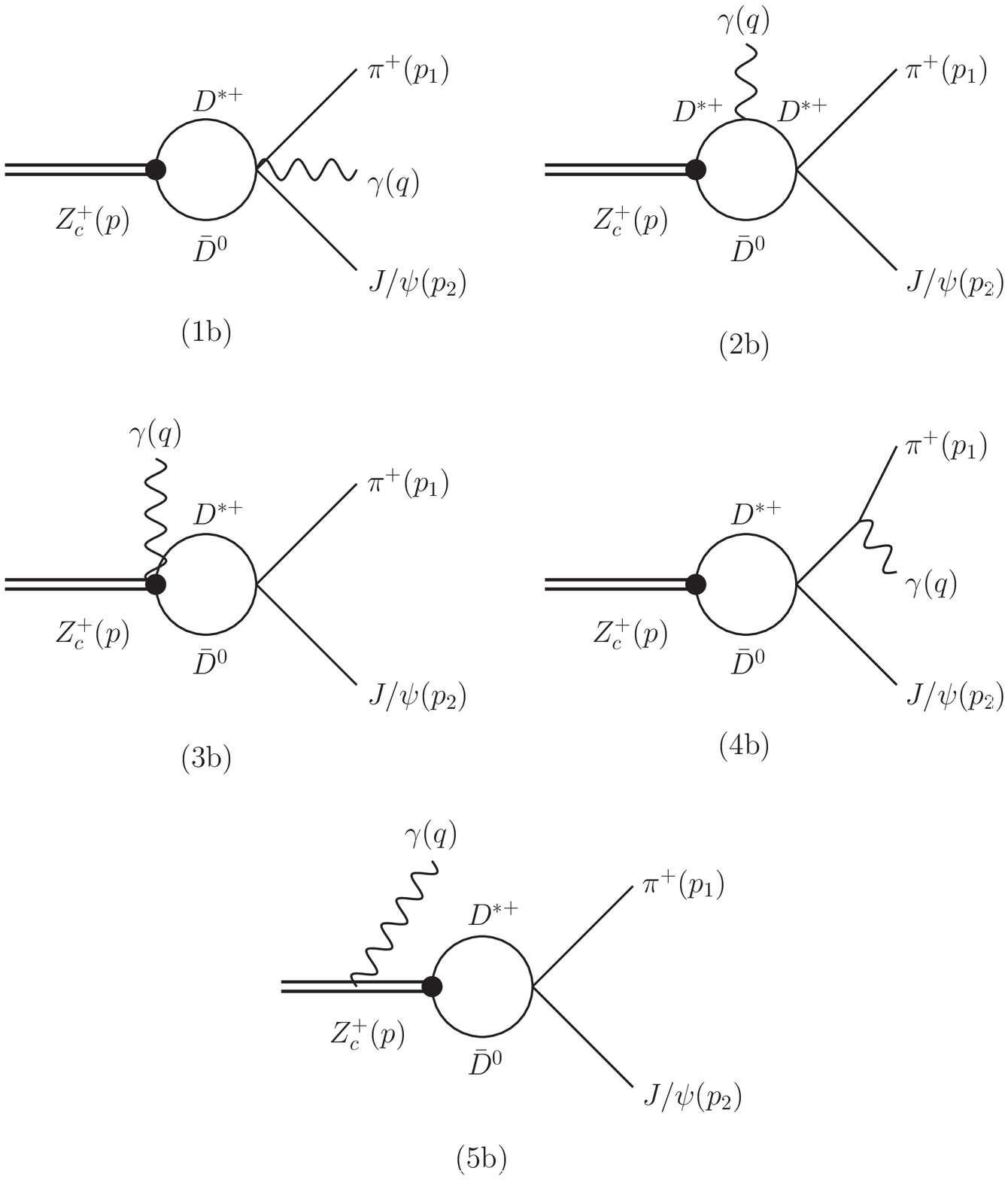}
	\caption{$\bar{D^{0}} D^{*+}$ meson loop diagrams 
        contributing to the three-body decay $\zc\to \jp\pi^+\gam$.}
	\label{diagramsZcJpsPiGamDsD2}
\end{figure}

The graphs governing the three-body decay $\zc\to \jp\pi^+\gam$ 
are shown in Figs.~\ref{diagramsZcJpsPiGamDsD1} 
and~\ref{diagramsZcJpsPiGamDsD2}.  
The diagrams are evaluated using the Schwinger representation 
for the propagators:
\eq
\frac{1}{m^2-k^2} = \int_{0}^{\infty} d\alpha \, e^{-\alpha(m^2-k^2)} \,.
\en
The resulting matrix element for the three-body decay $\zc\to \jp\pi^+\gam$ 
is gauge invariant as shown in Appendix~\ref{gaugeinv}. It can be decomposed 
into the following Lorentz structures: 
\eq
{\cal T}^{\alpha \beta \rho} &=& 
\sum\limits_{i=1}^5 {\cal T}^{\alpha \beta \rho}_i 
= \left( g^{\alpha \beta} p_2^{\rho} 
- p_2^{\alpha} g^{\beta \rho}\right ) \, 
F_1\nonumber\\ 
&+& \left(p_2 \cdot p_1 g^{\alpha \beta} 
- p_2^{\alpha} p_1^{\beta}\right ) p^{\rho} \, F_{235}
\nonumber\\
&+& \left((p-p_2)\cdot p_2 g^{\alpha \beta} 
- p_2^{\alpha} p^{\beta}\right) p_1^{\rho} \, F_4 
\en
where with $F_1$, $F_{235} = F_2 + F_3 + F_5$ and $F_4$ 
we denote structure integrals collected in Appendix~\ref{structint}.  
Since the diagrams $(2,3,5)$ in Fig. \ref{diagramsZcJpsPiGamDsD1} 
and Fig. \ref{diagramsZcJpsPiGamDsD2} have the same Lorentz structure 
they can be summed up together.

The matrix element ${\cal M}^{\alpha \beta}$ for the four-body decay 
$\translplm$ can be simply 
deduced from the matrix element for the three-body decay. 
We assume that only the photon contributes through conversion 
to the dilepton final state.
Hence the matrix element for the four-body decay factorizes into a 
three-body part ${\cal T}$ of the $\zc\to \jp\pi^+\gam$ transition 
and a leptonic part $\ell$ corresponding to the dilepton production 
$\gam\to \ell^+ \ell^-$:
\eq
{\cal M}^{\alpha \beta} 
= - \, \frac{e^2}{t} \ {\cal T}^{\alpha \beta \rho} \ \ell_{\rho}
\en
where $t = (k_3 + k_4)^2$, $\ell^{\rho} = \bar{u}(k_3) \, 
\gamma^{\rho} \, v(k_4)$ 
is the leptonic current and  $\bar{u}(k_3)$, 
$v(k_4)$ denote the spinors of the lepton and antilepton in the final 
state, respectively.

In the final state we sum over all polarizations.
The polarization sum factorizes 
into three different parts, one for the $Z_c$, 
one for the $\jp$, and one for the photon:
\eq 
\hspace*{-.5cm}
D^{\alpha_1\alpha_2}_{Z_c} &=& 
\sum\limits_{\rho = 1}^{3} \epsilon^{\alpha_1}_\rho(p) 
\epsilon^{\alpha_2}_\rho (p) \, = \, -g^{\alpha_1 \alpha_2}
+\frac{p^{\alpha_1}p^{\alpha_2}}{m_{Z_c}^2} \\
\hspace*{-.5cm}
D^{\beta_1\beta_2}_{\jp} &=&  
	\sum\limits_{\sigma = 1}^{3} \epsilon^{\beta_1}_\sigma(p_2) 
\epsilon^{\beta_2}_\sigma(p_2) \, = \, -g^{\beta_1 \beta_2}
+\frac{p_2^{\beta_1} p_2^{\beta_2}}{m_{\jp}^2} \\
\hspace*{-.5cm}
D^{\rho_1\rho_2}_{\gamma} &=& 
\sum\limits_{\lambda = 1}^{2} \epsilon^{\rho_1}_\lambda(q) 
\epsilon^{\rho_2}_\lambda (q) \, = \, - g^{\rho_1 \rho_2}
\en 
where $\epsilon^\alpha_\rho$ denotes the polarization vector. 
Using these, for the spin-averaged square of the 
$\zc\to \jp\pi^+\gam$ amplitude we write:
\eq 
\hspace*{-.5cm}	
\sum\limits_{\text{pol.}} \left|{\cal T}\right|^2 &=& 
D^{\rho_1\rho_2}_{\gamma}  \, 
D^{\alpha_1\alpha_2}_{Z_c} \, 
D^{\beta_1\beta_2}_{\jp} \,
{\cal T}_{\alpha_1 \beta_1 \rho_1} \, 
({\cal T}_{\alpha_2 \beta_2 \rho_2})^\dagger \, .
\en 
The leptonic $L^{\rho_1\rho_2}$ and hadronic $H^{\rho_1\rho_2}$ tensor  
contributing to the 
$\translplm$ decay rate are 
\eq 
\hspace*{-.5cm}
H_{\rho_1\rho_2} &=& D^{\beta_1\beta_2}_{J_\psi}  \, D^{\alpha_1\alpha_2}_{Z_c} \, {\cal T}_{\alpha_1 \beta_1 \rho_1} \, ({\cal T}_{\alpha_2 \beta_2 \rho_2})^\dagger \,,
\nonumber\\
\hspace*{-.5cm}
L^{\rho_1\rho_2} &=&
\sum\limits_{r,s} \bar{u}^{(r)}(k_3) 
\gamma^{\rho_1} v^{(s)}(k_4)\bar{v}^{(s)}(k_4) \gamma^{\rho_2} u^{(r)}(k_3) 
\nonumber\\
\hspace*{-.5cm}
&=& \text{tr}\left[(\cancel{k}_3+m)\gamma^{\rho_1}(\cancel{k}_4-m)
\gamma^{\rho_2}\right] 
\nonumber\\
\hspace*{-.5cm}
&=& 4 \, \left[ k_3^{\rho_1}k_4^{\rho_2}
+k_3^{\rho_2}k_4^{\rho_1}
-g^{\rho_1 \rho_2}(m_\ell^2+k_3k_4)
\right] \,, 
\en
where $m_\ell$ is the lepton mass. 
The spin-averaged square of the amplitude for the four-body decay  
$\translplm$ in terms of leptonic and hadronic tensors is finally written as 
\eq 
\sum\limits_{\text{pol.}} \left|{\cal M}\right|^2 = 
H^{\rho_1\rho_2} \, L_{\rho_1\rho_2} \,. 
\en
In the next step the invariant matrix element squared, 
$\sum_\text{pol.}\left|{\cal M}\right|^2$, will be expressed in terms 
of Lorentz scalar products of the five momenta $p_1, p_2, k_3, k_4$ and $p$. 
For the sake of simplicity we do not display the explicit, complicated result 
for $\sum_\text{pol.}\left|{\cal M}\right|^2$.

\section{Kinematics}
\label{kinematics}
\begin{figure}[ht]
	\centering
	\includegraphics[scale=1.25]{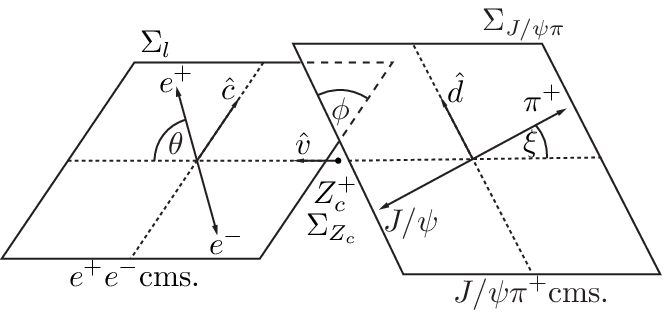}
	\caption{Choice of kinematical variables in the four-particle decay $\translplm$.}
	\label{fig:kinematics}
\end{figure}
To calculate the decay rates we have to specify two independent 
kinematical variables for the three-body decay $\zc\to \jp\pi^+\gam$ 
and five independent ones for the four-body decay  $\translplm$.
For the three-body decay we choose the invariant Mandelstam variables as 
\eq
	s_{12} &=& (p_1+p_2)^2\,,   \nonumber\\
	s_{31} &=& (p_1+q)^2  \,,   \nonumber\\
	s_{32} &=& (p_2+q)^2  \,,   \nonumber\\
s_{12} + s_{31} + s_{32} &=& m_{Z_c}^2 + m_\jp^2 + m_\pi^2 
+ \delta m_{\gamma}^2 \,, 
\en 
where 
$\delta m_{\gamma}^2$ is a cutoff parameter for 
the phase-space integration to avoid the infrared 
bremsstrahlung singularity
for $s_{31}\rightarrow m_{\pi}^2$ and 
$s_{12}\rightarrow m_{Z_c}^2$ 
[see Sec.~\ref{results} for further discussion].

With these definitions we can express the scalar products 
between the momenta $p_1, p_2$ and $q$ as 
\begin{align}
p_1 \cdot p_2 &= \frac{1}{2} (s_{12}-m_{\pi}^2-m_{\jp}^2) \,, \nonumber\\
p_1 \cdot q &= \frac{1}{2} (s_{31}-m_{\pi}^2-\delta m_{\gamma}^2)\,, \\ 
p_2 \cdot q &= \frac{1}{2} (m_{Z_c}^2+m_{\pi}^2-s_{12}-s_{31}) \,. \nonumber
\end{align}
The scalar products between $p$ and one outgoing momentum can be eliminated 
due to momentum conservation. Now the phase space region of the three-body 
decay can be expressed through the following ranges for the kinematical
variables \cite{Byckling:1973}:
\begin{align}
	(m_\pi+\delta m_{\gamma})^2 &\leq s_{31} \leq (m_{Z_c}-m_\jp)^2 \,, 
\nonumber\\
	s_{12}^- &\leq s_{12} \leq s_{12}^+ \,, 
\end{align}
where
\begin{align}
	s_{12}^\pm &= m_\jp^2+m_\pi^2 \nonumber\\
	&-\frac{1}{2 s_{31}}\left[(s_{31}-m_{Z_c}^2+m_\jp^2)(s_{31}+m_\pi^2
          -\delta m_{\gamma}^2\right.) \nonumber\\
	&\mp \left.\lambda^{1/2}(s_{31},m_{Z_c}^2,m_\jp^2)
        \lambda^{1/2}(s_{31},m_{\pi}^2,\delta m_{\gamma}^2)\right]
\end{align} 
and $\lambda(x,y,z)=x^2+y^2+z^2-2(xy+xz+yz)$ is the K\"allen triangle 
kinematical function.
\newline

To calculate the scalar products between the momentum vectors $p_1, p_2, k_3$ 
and $k_4$ we will consider three reference frames for the four particle phase 
space: the rest frame $\Sigma_{Z_c}$ of the $Z_c$ meson, the dilepton 
center-of-mass frame $\Sigma_l$, and the center-of-mass frame
$\Sigma_{J/\psi\pi}$ of the $\jp~\pi$-pair [see Fig.~\ref{fig:kinematics}].   
For the four-body decay we choose the kinematical variables 
suggested in Ref.~\cite{Cabibbo:1965zzb} and extensively used e.g. 
in Refs.~\cite{Bijnens:1992mk}-\cite{Knochlein:1996ah}:   
\begin{itemize}[noitemsep,nolistsep]
	\item $s_{12}$, the invariant mass squared of the $\jp~\pi$-pair
	\item $s_{34}$, the invariant mass squared of the lepton pair
	\item  $\theta$, the angle of the antilepton $l^+$ in $\Sigma_l$ 
with momentum $k_3$ and with respect to the dilepton line of flight 
in $\Sigma_{Z_c}$
	\item $\xi$, the angle of the pion $\pi^+$ in $\Sigma_{J/\psi\pi}$ 
with momentum $p_1$ and with respect to the dimeson line of flight in 
$\Sigma_{Z_c}$ 
	\item $\phi$, the angle between the plane formed by the mesons 
$\jp$, $\pi$ in $\Sigma_{Z_c}$ and the corresponding plane formed by the 
dileptons
\end{itemize}
It proves to be very helpful to introduce linear combinations 
of the momenta $p, p_1, p_2, k_3$ and $k_4$. One of these momenta 
can always be eliminated due to momentum conservation:
\begin{align*}
	K &= p_1 + p_2\,, &L = p_1 - p_2\,, \\
	Q &= k_3 + k_4\,, &R = k_3 - k_4\,.
\end{align*}
In order to express the Lorentz scalar products in terms of 
the kinematical variables specified above, we need the following 
expressions with the general masses $M, m_1, m_2, m_3, m_4$ 
(note that these expressions will hold for any four-body decay 
with the frames specified in Fig.~\ref{fig:kinematics})
\begin{align}
	K\cdot K &= s_{12}\,, \nonumber\\
	Q\cdot Q& = s_{34}\,, \nonumber\\
	K\cdot Q &= \frac{1}{2} (M^2-s_{12}-s_{34})\,, \nonumber\\
	K \cdot L &= m_1^2-m_2^2\,, \nonumber\\
	Q \cdot R &= m_3^2 - m_4^2\,, \nonumber\\
	Q \cdot L &= \frac{(K \cdot L)}{(K \cdot K)} 
        (K \cdot Q) + x \sigma_{12} \cos\xi\,, \\
	K \cdot R &= \frac{(Q \cdot R)}{(Q \cdot Q)} 
        (K \cdot Q) + x \sigma_{34} \cos\theta\,, \nonumber\\
	R \cdot L &= \frac{(Q \cdot R)}{(Q \cdot Q)}    
        x \sigma_{12} \cos\xi +\frac{(K \cdot L)}{(K \cdot K)} 
        x \sigma_{34} \cos\theta\,\nonumber\\
        &+(K\cdot Q) \sigma_{12}\sigma_{34} \cos\xi\cos\theta 
        + \frac{(Q\cdot R)}{(Q \cdot Q)}\frac{(K \cdot L)}
        {(K \cdot K)} (K\cdot Q) 
        \nonumber\\
	&-\sqrt{s_{12}s_{34}}\sigma_{12}\sigma_{34}\sin\xi\sin\theta\cos\phi 
\nonumber\\
\sigma_{ij} &=  \frac{\lambda^{1/2}(s_{ij},m_i^2,m_j^2)}{s_{ij}}, \quad 
x \, = \, \frac{1}{2}\lambda^{1/2}(M^2,s_{12},s_{34})  \,. \nonumber
\end{align}
These relations are obtained by calculating the Lorentz boosts between the
different frames $\Sigma_l$, $\Sigma_{J/\psi\pi}$ and $\Sigma_{Z_c}$ as done 
in~\cite{Daphne:1992}.
Now all scalar products between the outgoing momenta occurring in the 
spin-averaged amplitude $\sum_\text{pol.}\left|{\cal M}\right|^2$ for 
the four-body decay can be written as
\eq
	p_1 \cdot p_2 &=& \frac{1}{2}(s_{12}-m_1^2-m_2^2)\,, \nonumber\\
	k_3 \cdot k_4 &=& \frac{1}{2}(s_{34}-m_3^2-m_4^2)\,, \nonumber\\
	p_1 \cdot k_3 &=& \frac{1}{4} (Q\cdot K + Q \cdot L+R\cdot K+R\cdot L)
\,, \nonumber\\
	p_2 \cdot k_3 &=& \frac{1}{4} (Q\cdot K - Q \cdot L+R\cdot K-R\cdot L)
\,, \\
	p_1 \cdot k_4 &=& \frac{1}{4} (Q\cdot K + Q \cdot L-R\cdot K-R\cdot L) 
\,, \nonumber\\
	p_2 \cdot k_4 &=& \frac{1}{4} (Q\cdot K - Q \cdot L-R\cdot K+R\cdot L)
\,.  
\nonumber 
\en
The ranges of the variables, which define the limits 
of the phase-space integration, are
\eq 
	(m_3+m_4)^2 &\leq& s_{34} \ \leq \ (M-m_1-m_2)^2\,, \nonumber\\
	(m_1+m_2)^2 &\leq& s_{12} \ \leq \ (M-\sqrt{s_{34}})^2\,, \\
	0 &\leq& \theta, \xi \ \leq \ \pi\,, \nonumber\\
	-\pi &\le& \phi \ \leq \ \pi \,. \nonumber
\en
In our case we set $M = m_{Z_c}$, $m_1 = m_{\pi}$, $m_2 = m_{\jp}$, 
and $m_3 = m_4 = m_l$.
The decay rates can then be written as (see, e.g.,~\cite{Byckling:1973})
\eq
& &\Gamma_3(\zc\to \jp\pi^+\gam) = 
\frac{1}{N_3} \int ds_{31} \int ds_{12} 
\sum_{\text{pol.}} \left| {\cal T}\right|^2 \,, \nonumber\\
& &\\
& &\Gamma_4(\zc\to \jp\pi^+\ell^+\ell^-) =   
\frac{1}{N_4} \int ds_{34} \int ds_{12} \nonumber\\
&\times&\int d\cos\theta\, d\cos\xi\, d\phi\, x s\sigma_{12} \sigma_{34}
\sum_{\text{pol.}} \left| {\cal M}\right|^2 \,, \nonumber
\en 
where $N_3 = 3 \times 2^8 \pi^3 m_{Z_c}^3$ and 
$N_4 = 3 \times 2^{15} \pi^6 m_{Z_c}^3$. 

\section{Numerical results}
\label{results}

With the phenomenological Lagrangians, kinematics and partial evaluation
of the transition amplitudes introduced 
we now can proceed to determine the widths of the
three- and four-body decays numerically. 

\begin{figure}[hb]
	\centering
	\includegraphics[scale=.55]{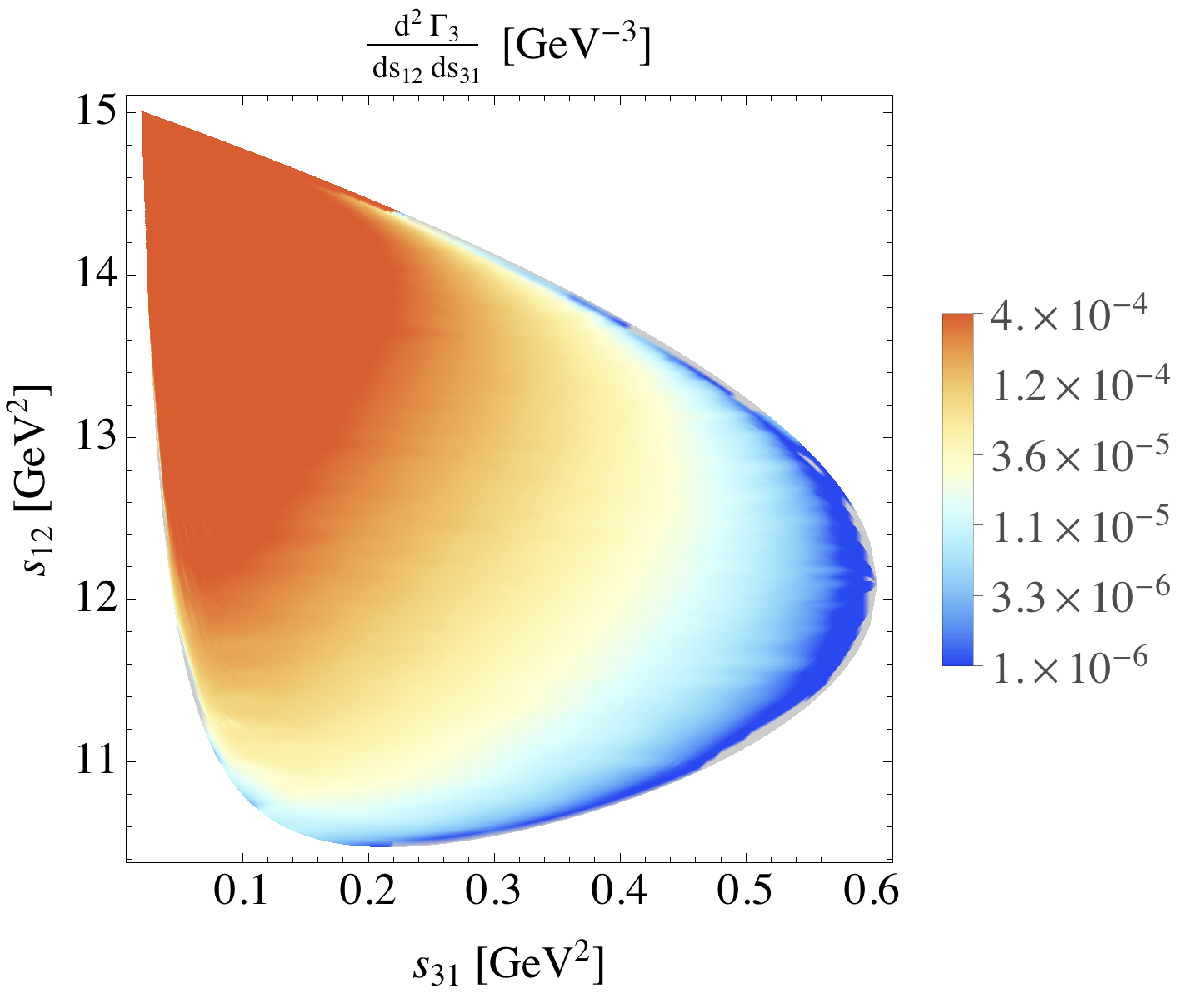}
	\caption{Double differential decay rate $d^2\Gamma_3/ds_{12} ds_{31}$ 
in $\unit{GeV^{-3}}$ for the three-body decay $\zc\to \jp\pi^+\gamma$
($\epsilon = 3 \unit{MeV}$ and $\Lambda = 650 \unit{MeV}$).}
	\label{fig:dGammads12ds31ZcJPsiPiGam1}
	\centering
	\includegraphics[scale=0.55]{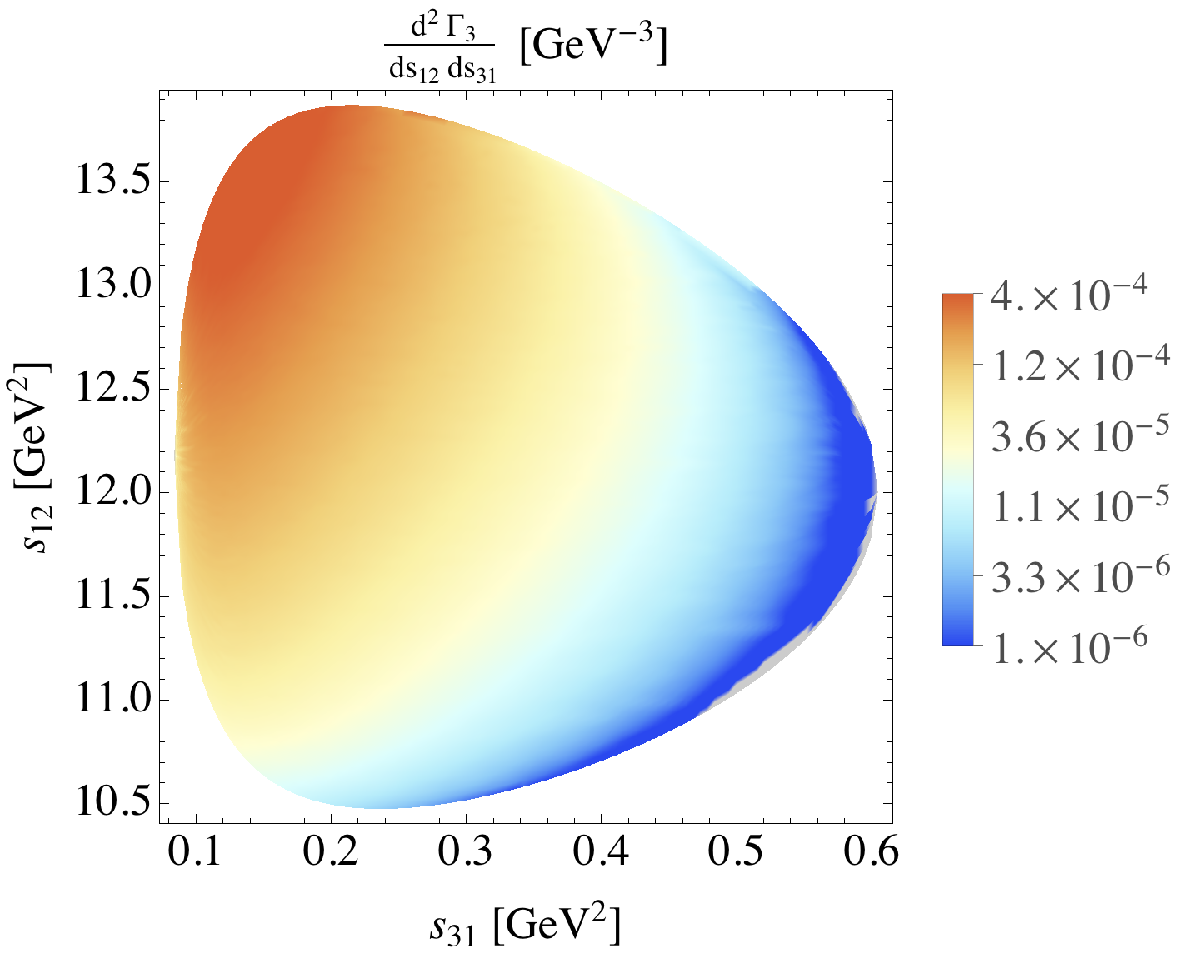}
	\caption{Double differential decay rate $d^2\Gamma_3/ds_{12} 
ds_{31}$ in 
$\unit{GeV^{-3}}$ for the three-body decay $\zc\to \jp\pi^+\gamma$
($\epsilon = 3 \unit{MeV}$ and $\Lambda = 650 \unit{MeV}$). 
The contour is nearly the same as in Fig.~\ref{fig:dGammads12ds31ZcJPsiPiGam1} 
except for $s_{31}\rightarrow m_{\pi}^2$ and $s_{12}\rightarrow m_{Z_c}^2$ 
where the bremsstrahlung singularity is located. 
This area is now excluded with an energy cut 
at $\delta m_\gamma = 150 \unit{MeV}$.}
	\label{fig:dGammads12ds31ZcJPsiPiGam3}
\end{figure}


\begin{figure}[hb]
	\centering
	\includegraphics[scale=0.55]{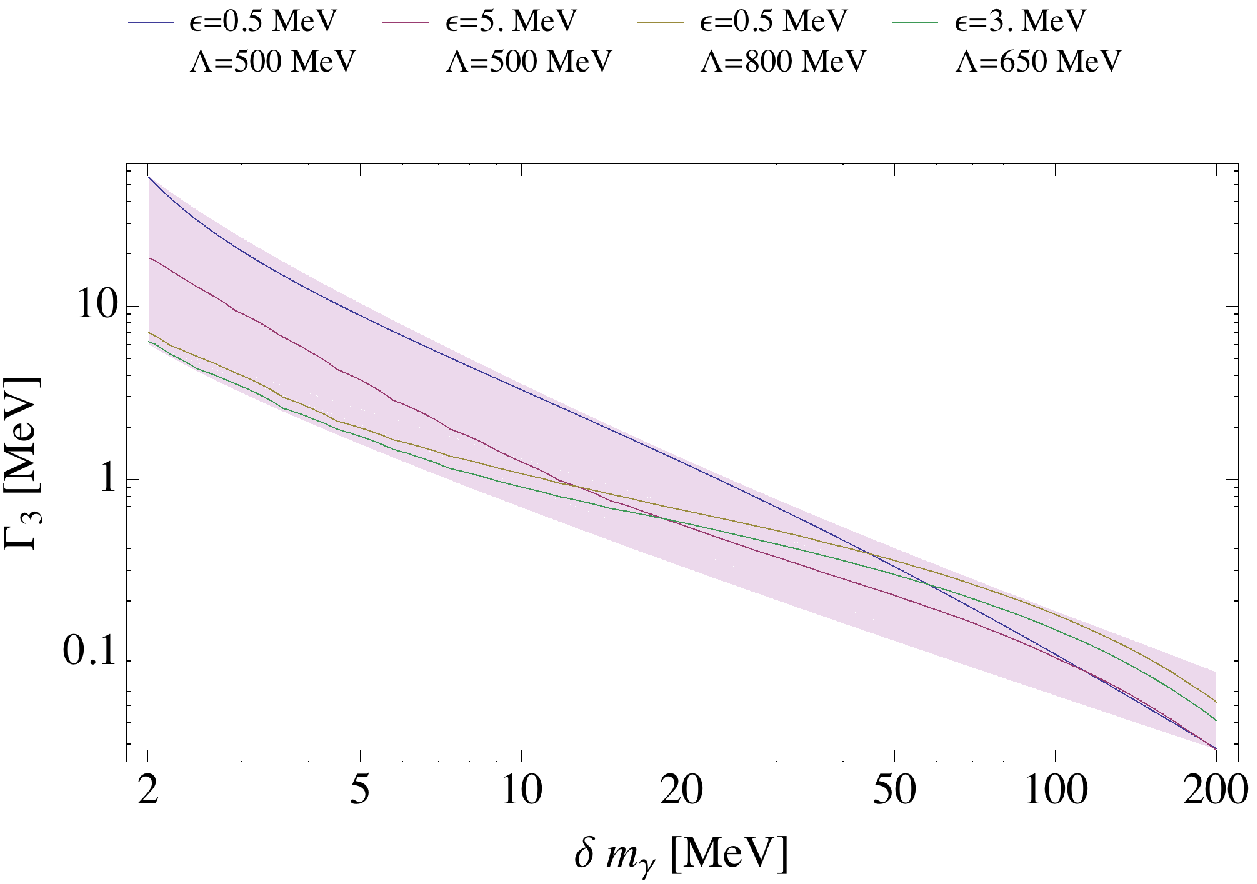}
	\caption{Partial decay rate $\Gamma_3$ in $\unit{MeV^{-1}}$ 
as a function of $\delta m_\gamma$ for selected values of $\epsilon$ 
and $\Lambda$ for the three-body decay $\zc\to \jp\pi^+\gamma$.}
	\label{fig:GammaMgam}
	\centering
	\includegraphics[scale=0.55]{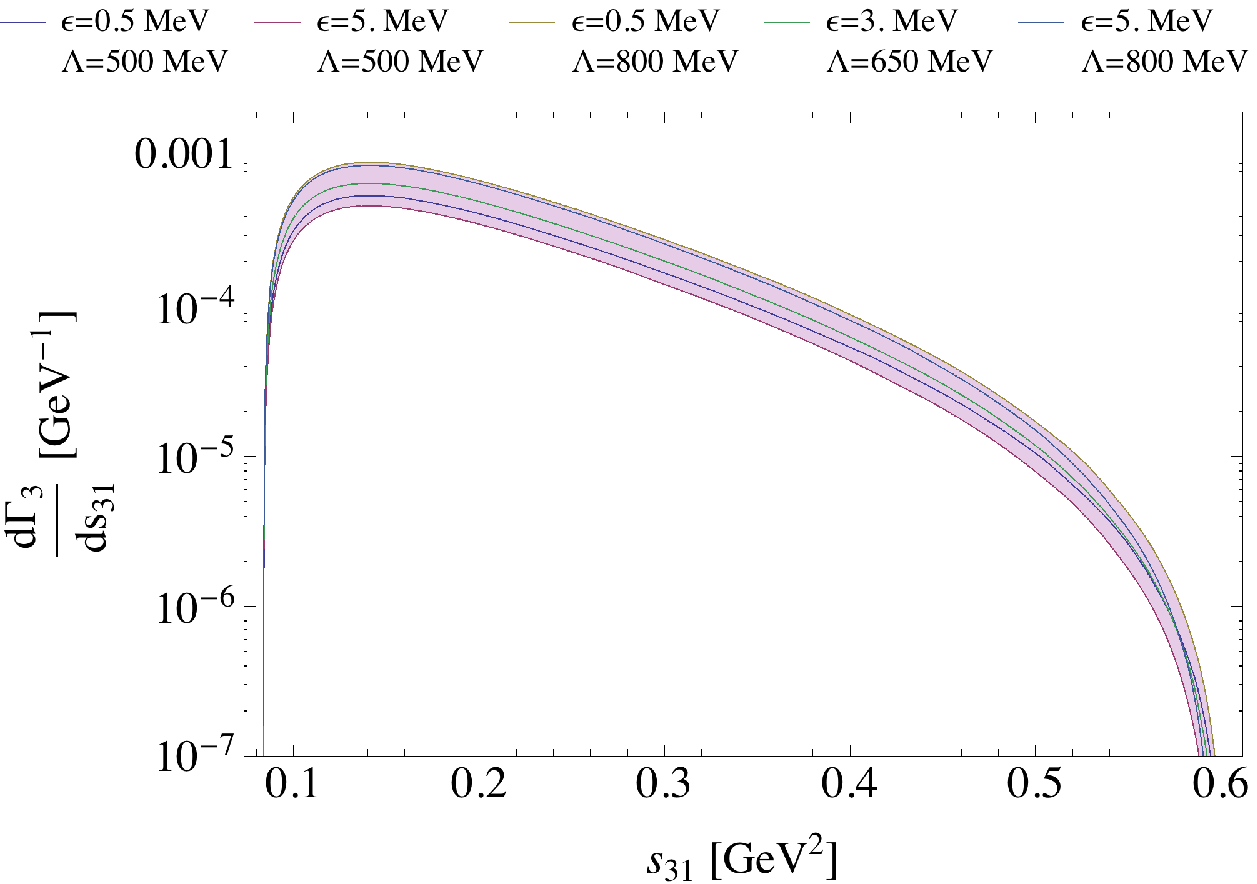}
	\caption{Differential decay rate $d\Gamma_3/ds_{31}$ 
in $\unit{GeV^{-1}}$ 
for selected values of $\epsilon$ and $\Lambda$ for the three-body decay 
$\zc\to \jp\pi^+\gamma$ with $\delta m_\gamma = 150\unit{MeV}$.}
	\label{fig:dGammads31ZcJPsiPiGam}
	\centering
	\includegraphics[scale=0.55]{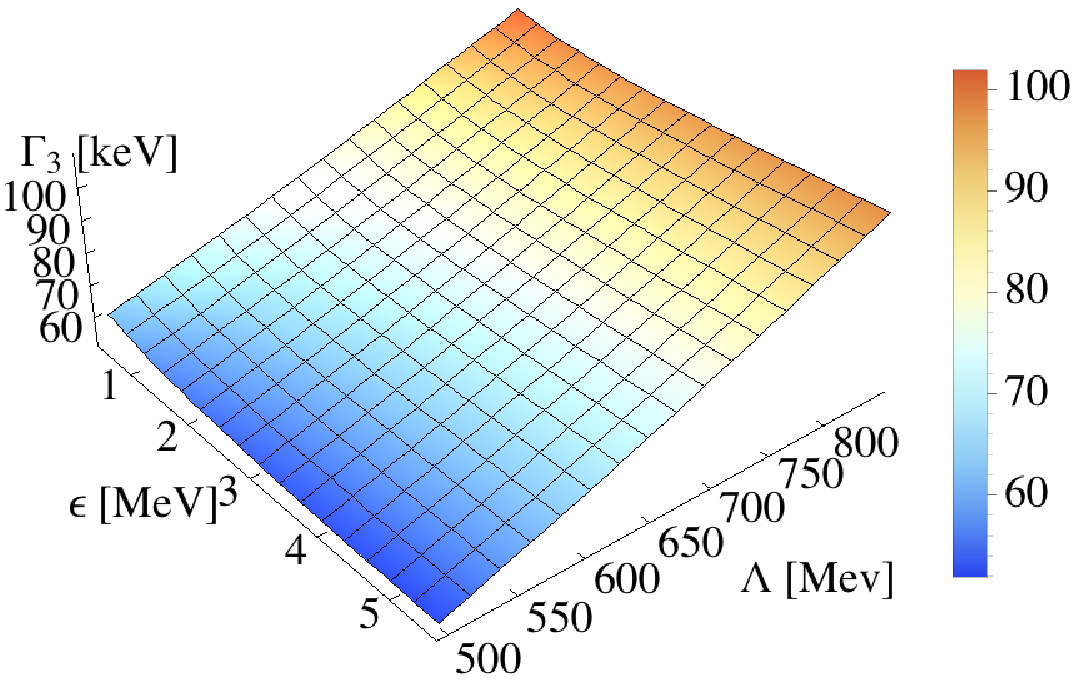}
	\caption{Partial decay rate $\Gamma_3$ in $\unit{keV}$ in dependence 
on $\epsilon$ and $\Lambda$ for the three-body decay $\zc\to \jp\pi^+\gamma$ 
with $\delta m_\gamma = 150\unit{MeV}$.}
	\label{fig:GammaZcJPsiPiGam}
\end{figure}

A full list of the results for the coupling $g_{Z_c}$ and the
decay widths is tabulated in Appendix~\ref{AppDecay}. Table I
contains the values for the coupling $g_{Z_c}$. 
In Tables II-IV we display the predictions 
for the three- and four-body decay rates for different values 
of the binding energy with $\epsilon = 0.5 \unit{MeV}$ to $5 \unit{MeV}$
and of the cutoff in the vertex function with
$\Lambda = 500 \unit{MeV}$ to $800 \unit{MeV}$. 
A substantial increase of the size parameter $\Lambda$, 
more suitable for a compact bound state, would lead to 
a sizable increase in the decay rates. 
Therefore, if experiment will deliver larger values for 
the decay rates than predicted in our approach, 
this could signal that the $Z_c^+$ is probably not a molecular state.

\begin{figure}[htb]
	\centering
	\includegraphics[scale=0.535]{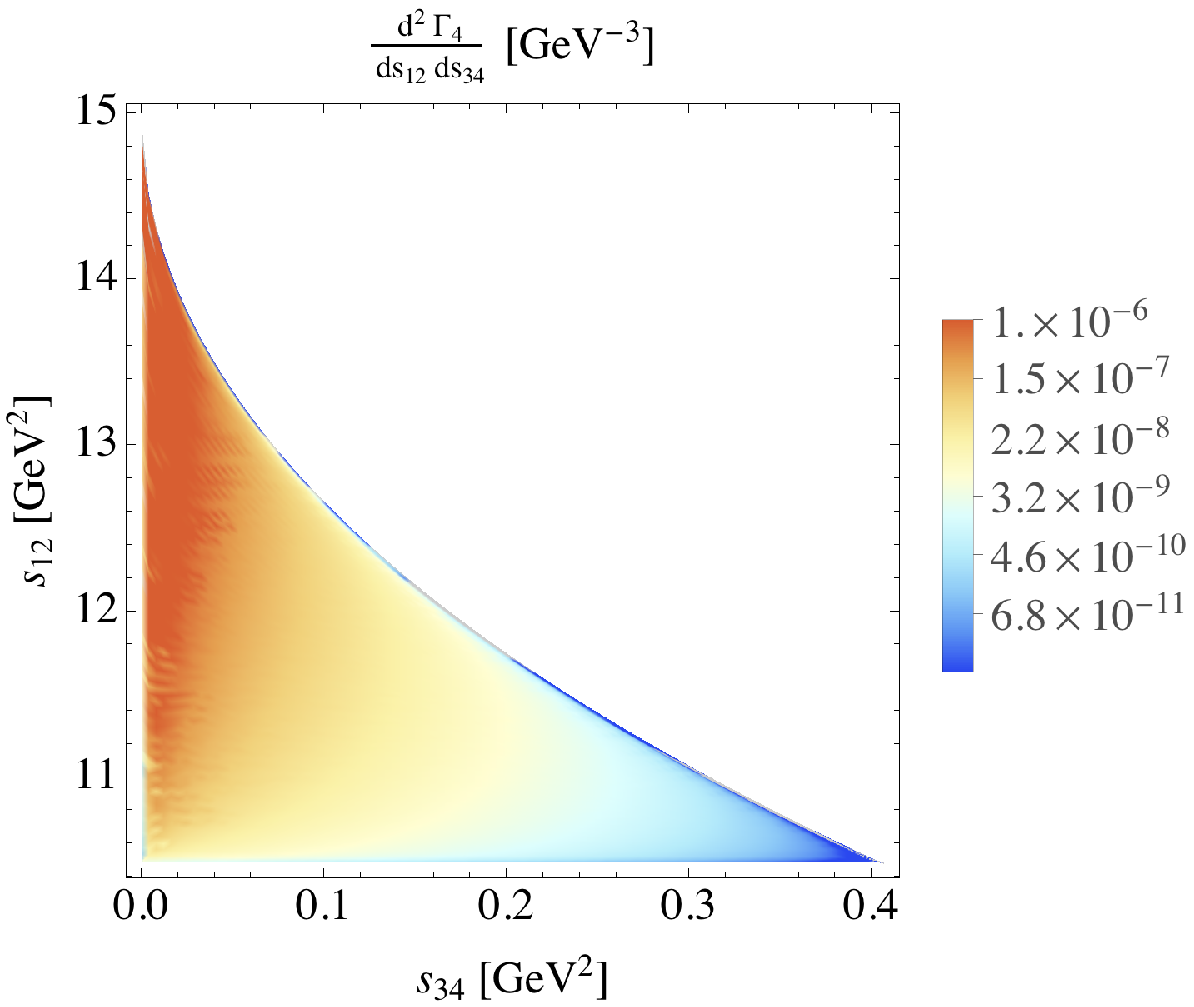}
	\caption{Double differential decay rate $d^2\Gamma_4/ds_{12} ds_{34}$ 
in $\unit{GeV^{-3}}$ for $\epsilon = 3\unit{MeV}$ and 
$\Lambda = 650\unit{MeV}$ for the four-body decay $\zc\to \jp\pi^+\ep\em$.}
	\label{fig:dGammads12ds34ZcJPsiPiepem2}
	\centering
	\includegraphics[scale=0.535]{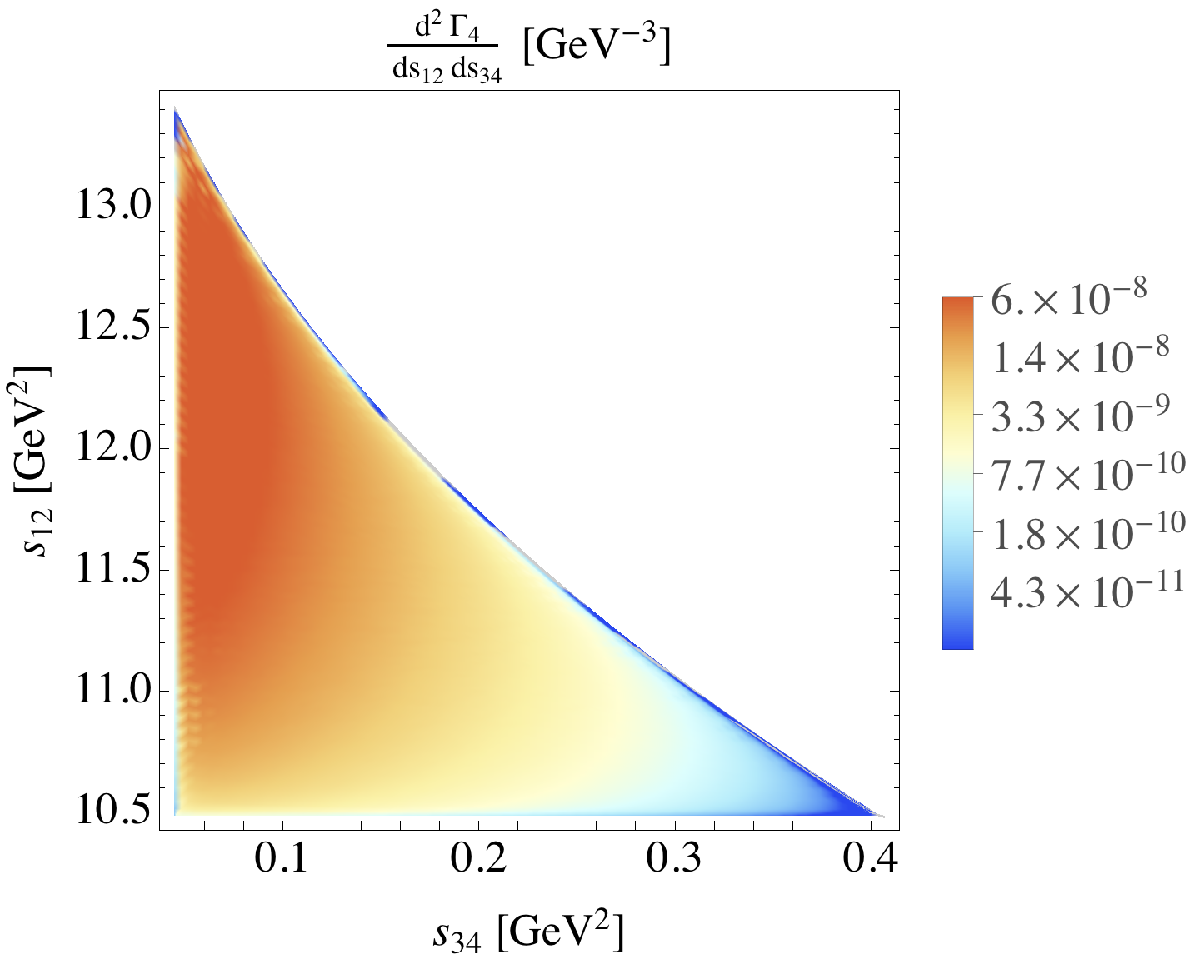}
	\caption{Double differential decay rate $d^2\Gamma_4/ds_{12} ds_{34}$ 
in $\unit{GeV^{-3}}$ for 
$\epsilon = 3 \unit{MeV}$ and $\Lambda = 650 \unit{MeV}$ 
for the four-body decay $\zc\to \jp\pi^+\mup\mum$.}
	\label{fig:dGammads12ds34ZcJPsiPimupmum2}
\end{figure}

Let us first discuss the Dalitz plot $d^2\Gamma/ds_{12}ds_{31}$ for 
the three-body transition $\transGam$. The contour plot in 
Fig.~\ref{fig:dGammads12ds31ZcJPsiPiGam1} shows an infrared bremsstrahlung 
singularity for the limits $s_{31}\rightarrow m_{\pi}^2$ and 
$s_{12}\rightarrow m_{Z_c}^2$. 
For these values of $s_{31}$ and $s_{12}$ the bremsstrahlung diagrams 
$4$a, $4$b, $5$a and $5$b in Figs. \ref{diagramsZcJpsPiGamDsD1} 
and~\ref{diagramsZcJpsPiGamDsD2} will generate a divergence 
in the double differential decay rate $d^2\Gamma/ds_{12}ds_{31}$.

A measurement of the partial or differential decay rate depends on the minimum 
photon energy detectable in the experimental facility. 
Hence, both in experiment 
and in theory, it is only possible to determine the partial or 
differential decay 
rate as a function of an energy cut $\delta m_\gamma$. To handle the 
bremsstrahlung singularity for $s_{31}\rightarrow m_{\pi}^2$ and 
$s_{12}\rightarrow m_{Z_c}^2$ we will use an energy cut at 
$\delta m_\gamma = 150 \unit{MeV}$ which holds for most facilities.
When we apply the energy cut to the Dalitz plot in 
Fig.~\ref{fig:dGammads12ds31ZcJPsiPiGam1} 
we obtain Fig.~\ref{fig:dGammads12ds31ZcJPsiPiGam3} which has 
no bremsstrahlung singularity any more. 
In Fig.~\ref{fig:GammaMgam} we show the partial decay rate 
for the three-body transition $\transGam$ as a function of $\delta m_\gamma$.
In Fig.~\ref{fig:dGammads31ZcJPsiPiGam} we give the differential decay rate 
$d\Gamma_3/ds_{31}$ which has been evaluated from $d^2\Gamma_3/ds_{12}ds_{31}$ 
by integration over $s_{12}$ for an energy cut at 
$\delta m_\gamma = 150\unit{MeV}$.

\begin{figure}
	\centering
	\includegraphics[scale=0.55]{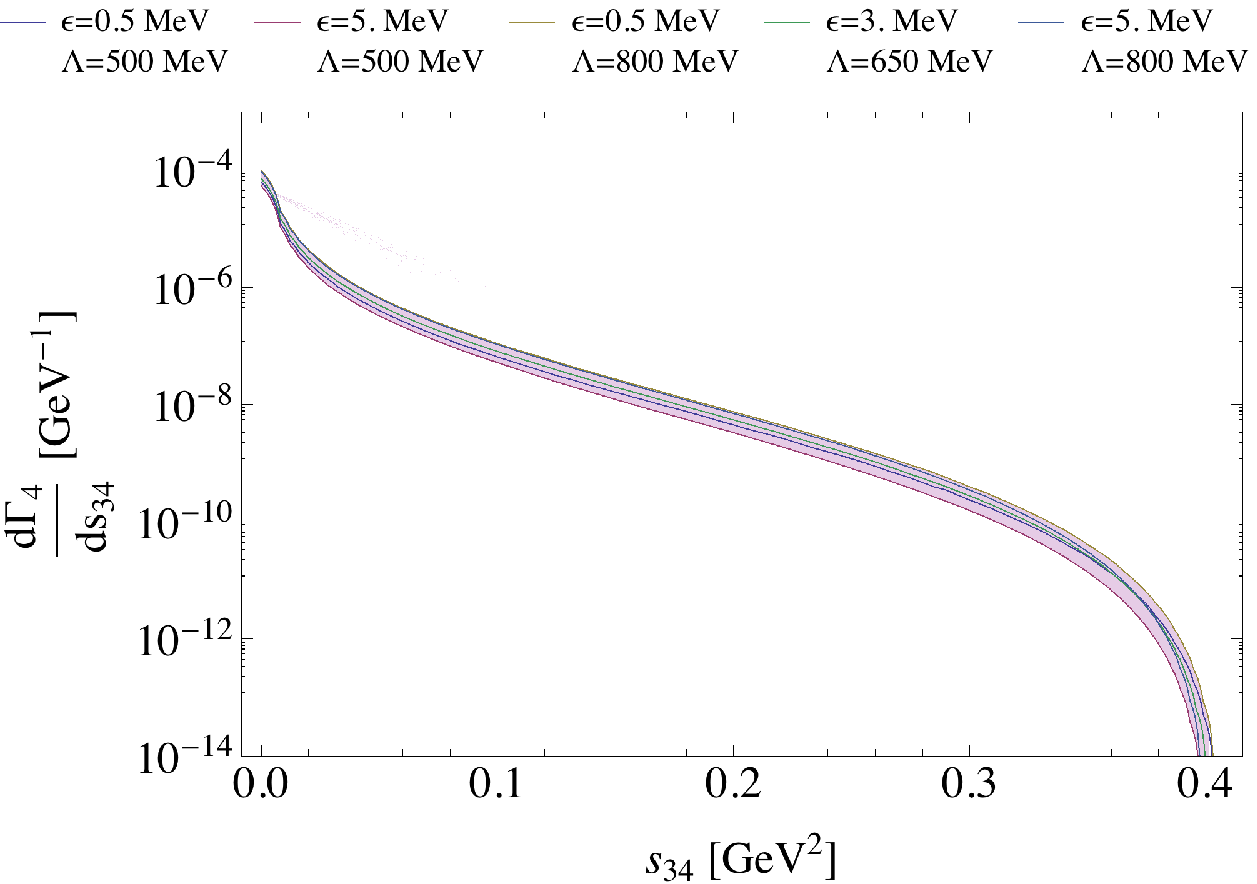}
	\caption{Differential decay rate $d\Gamma_4/ds_{34}$ 
in $\unit{GeV^{-1}}$ 
for selected values of $\epsilon$ and $\Lambda$ for the four-body decay 
$\zc\to \jp\pi^+\ep\em$.}
	\label{fig:dGammads34ZcJPsiPiepem3}
	\centering
	\includegraphics[scale=0.55]{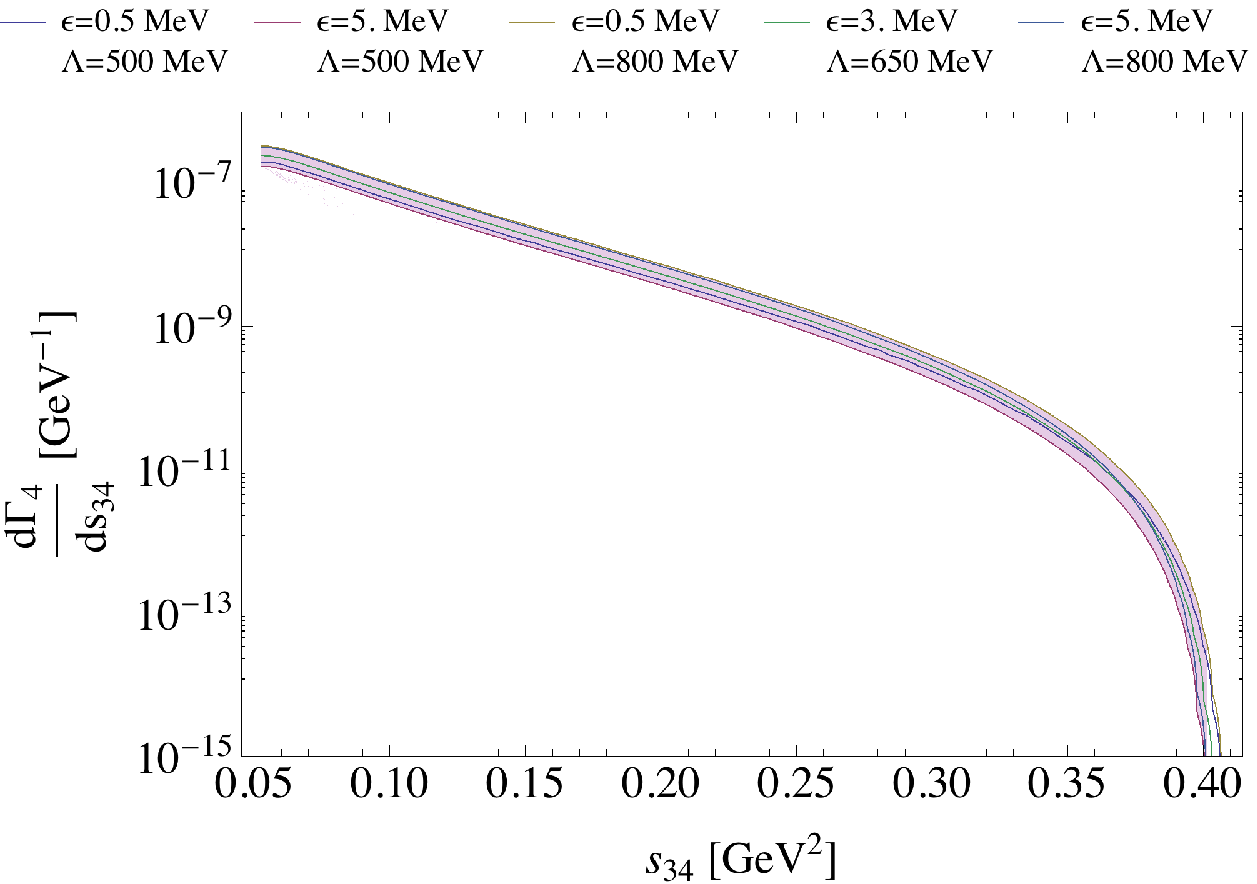}
	\caption{Differential decay rate $d\Gamma_4/ds_{34}$ in $\unit{GeV^{-1}}$ 
for selected values of $\epsilon$ and $\Lambda$ for the four-body decay 
$\zc\to \jp\pi^+\mup\mum$.}
	\label{fig:dGammads34ZcJPsiPimupmum3}
	\centering
	\includegraphics[scale=0.55]{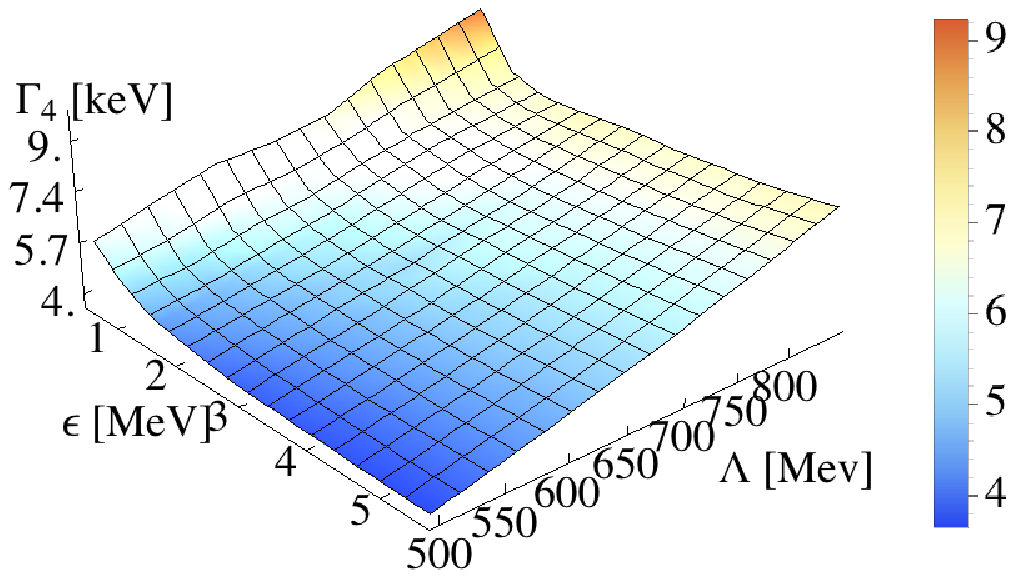}
	\caption{Partial decay rate $\Gamma_4$ in $\unit{keV}$ as a function of 
$\epsilon$ and $\Lambda$ for the four-body decay $\zc\to \jp\pi^+\ep\em$.}
	\label{fig:GammaZcJPsiPiepem}
	\centering
	\includegraphics[scale=0.55]{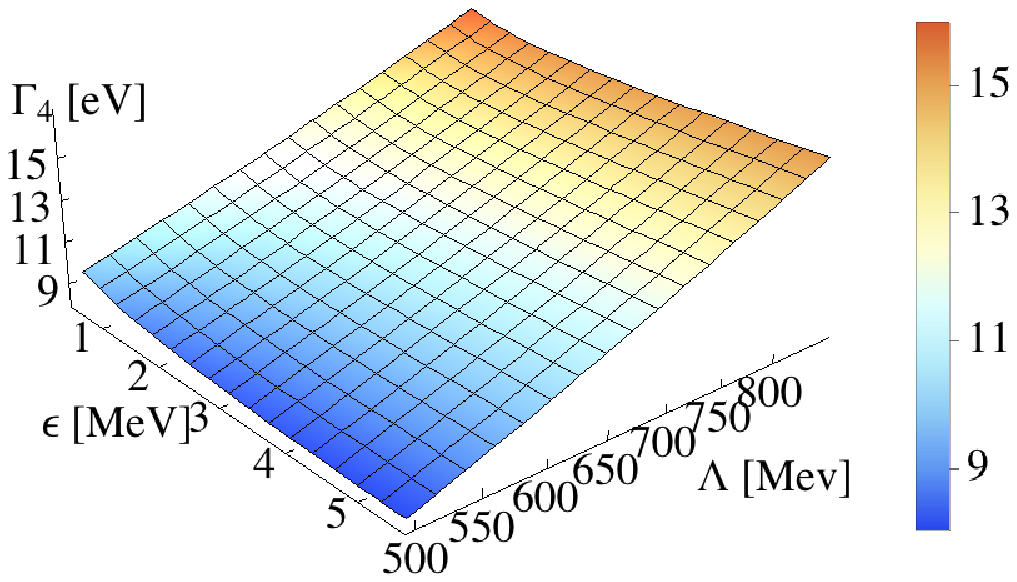}
	\caption{Partial decay rate $\Gamma_4$ in $\unit{keV}$ as a function of 
$\epsilon$ and $\Lambda$ for the four-body decay $\zc\to \jp\pi^+\mup\mum$.}
	\label{fig:GammaZcJPsiPimupmum}
\end{figure}

In Fig.~\ref{fig:GammaZcJPsiPiGam} we demonstrate the sensitivity of the decay 
rate $\Gamma_3$ on variations of the free parameters $\epsilon$ and $\Lambda$ for 
an energy cut at $\delta m_\gamma = 150 \unit{MeV}$. We note that the dependence 
is rather flat, the decay rate does not change significantly under the considered 
variations of 
$\epsilon$ and $\Lambda$.

To avoid the bremsstrahlung singularity for $s_{31}\rightarrow m_{\pi}^2$ and 
$s_{12}\rightarrow m_{Z_c}^2$, another physics possibility is available: 
a lepton-antilepton pair can be attached to the photon line as described in Sec. \ref{framework}.
Although the phase-space treatment for the four-body decays $\transepem$ and $\transmupmum$ 
gets more complicated, now an energy cut $\delta m_\gamma$ is not needed, the differential 
decay rates do not diverge.
For the typical values of $\epsilon = 3\unit{MeV}$ and $\Lambda = 650\unit{MeV}$ 
we obtain the Dalitz plots $d^2\Gamma_4/ds_{12}ds_{34}$ depicted in 
Figs.~\ref{fig:dGammads12ds34ZcJPsiPiepem2} and \ref{fig:dGammads12ds34ZcJPsiPimupmum2} 
for the four-body decays $\transepem$ and $\transmupmum$.

The differential decay rates $d\Gamma_4/ds_{34}$, which have been evaluated from 
$d^2\Gamma_4/ds_{12}ds_{34}$ by integration over $s_{12}$, are displayed in 
Figs.~\ref{fig:dGammads34ZcJPsiPiepem3} and \ref{fig:dGammads34ZcJPsiPimupmum3}.
We demonstrate the sensitivity of the decay rates $\Gamma_4$ on variations of 
the free parameters $\epsilon$ and $\Lambda$ in Figs.~\ref{fig:GammaZcJPsiPiepem} 
and \ref{fig:GammaZcJPsiPimupmum}. Here again, the decay rates do not change 
significantly under variations of $\epsilon$ and~$\Lambda$.

\section{Summary}

We have discussed the three- and four-body decays $\transGam$ and $\translplm$ of 
the $\zc(3900)$ considered as a hadronic $\bar{D}D^*$ molecule in a phenomenological 
Lagrangian approach. Our approach is manifestly Lorentz and gauge invariant and 
is based on the use of the compositeness condition. We have only two model parameters: 
the binding energy $\epsilon$ and $\Lambda$, which is related to the size of the 
$\bar{D}D^*$ distribution in the $\zc$-meson and, therefore, controls finite-size 
effects. The detailed results given for these decays are typical for a molecular state.
A naive estimate for a compact configuration would correspond to considerably larger values
of $\Lambda$ leading to a sizable enhancement of these decay rates.
But this effect should be confirmed by an explicit calculation of the decay
modes for a tetraquark interpretation of the $\zc$.

To summarize, when interpreting the $\zc$ as a $\bar{D}D^*$ molecule the resulting values
for the decay widths are $50$-$100\unit{keV}$ for the transition $\transGam$,
$4$-$10\unit{keV}$ for $\transepem$ and $8$-$16\unit{eV}$ for the 
decay $\transmupmum$.  
The predictions given here can add to the understanding of the $\zc$ structure
once the decay modes become accessible experimentally.
 
To elaborate further on a possible molecular structure of the $\zc(3900)$ 
in future we plan to examine the transition $\transGamp$ and $\translplmp$ for 
the possible partner state $\zcp(3900)$, which is treated as a $\bar{D^*}D^*$ 
molecule~\cite{Dong:2013iqa}.
As another possible continuation of this work $\zc$ decays can also be 
studied in the tetraquark model~\cite{Dubnicka:2010kz}. This approach 
is also based on the compositeness condition and was successfully 
applied to the study of the $X(3872)$ as a possible tetraquark state.
A full treatment of these observables for various structure interpretations
can possibly help to understand the nature of these unusual meson states.

\begin{acknowledgments}

This work is supported by the DFG under Contract No. LY 114/2-1 and
by Tomsk State University Competitiveness Improvement Program.

\end{acknowledgments}

\appendix
\appendix
\section{Mass operator and coupling constant} 
\label{massop}
The expressions for the coupling constant $g_{_{Z_c}}$ is
\eq 
	g_{Z_c}^{-2} &=& \Sigma_\perp^\prime(m_{Z_c}^2)  
= \frac{m_{Z_c}^2}{32 \pi^2} \int_{0}^{\infty} dx dy 
\frac{l}{a^3} \left( 1+ \frac{1}{2 m_{D^\ast}^2 a} \right) \nonumber\\
&\times&\exp\left( m_{Z_c}^2 \frac{l}{2 a} - m_D^2 x - m_{D^\ast}^2 y 
\right)\,, 
\en 
where $a =  2s + x + y, l =  s x + s y + 2 x y, s = \Lambda^{-2}$.

The numerical values are given in Table~I.

\section{Feynman rules} 
\label{feynrules}
\begin{figure}[htb]
	\centering
	\includegraphics[width=1\linewidth]{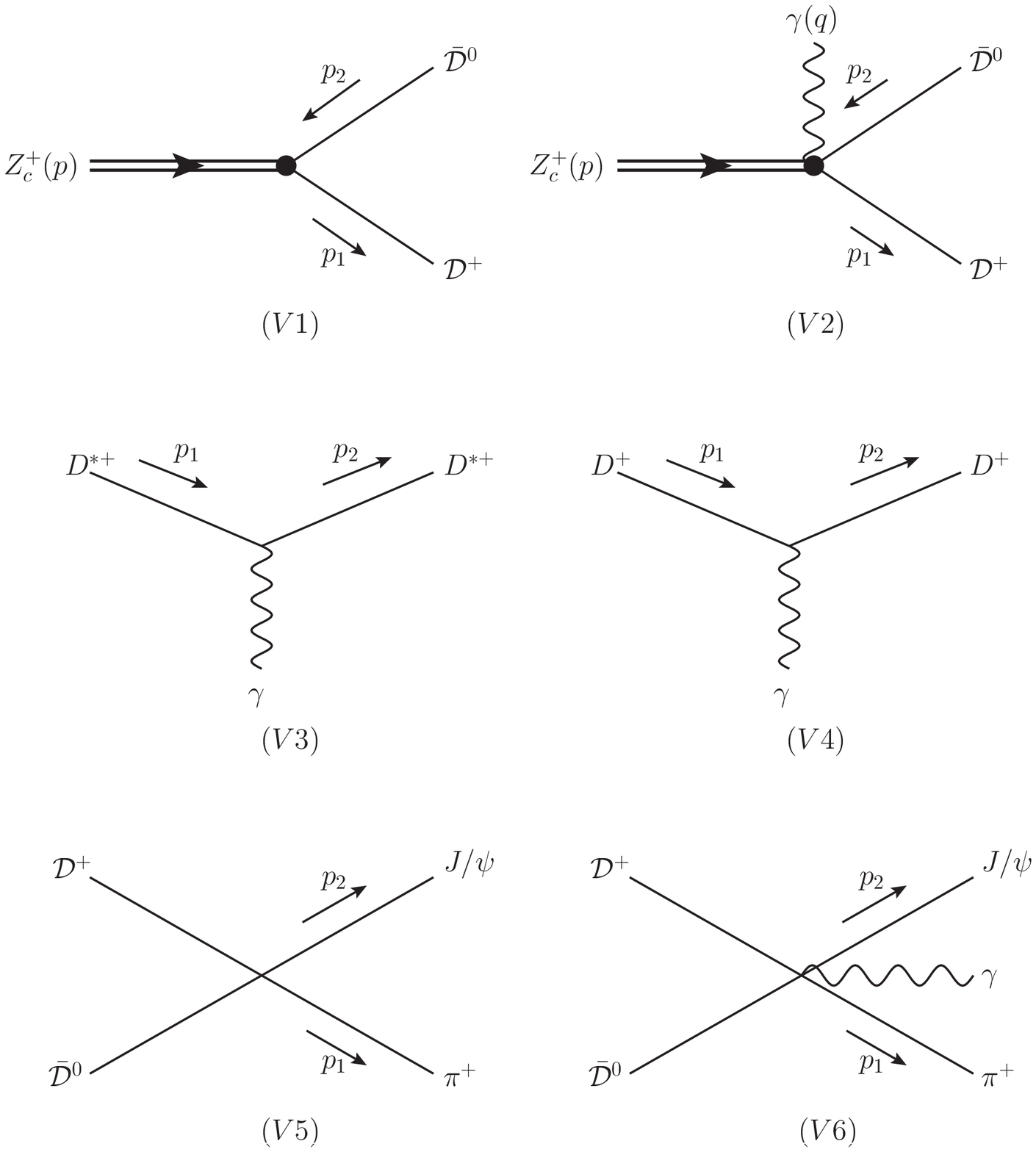}
	\caption{Vertices contributing to the three-body decay 
 $\zc\to \jp\pi^+\gam$. 
}
	\label{diagramsVertices}
\end{figure}

Since nonlocal gauge theories are not so common, 
we will briefly indicate the relevant Feynman rules in this Appendix. 
In our calculations we use
\eq 
D^{\mu \nu} (k) &=& \frac{i\biggl(-g^{\mu \nu}
+\frac{k^\mu k^\nu}{M_V^2}\biggr)}{k^2-M_V^2} 
\ \approx \ -\frac{ig^{\mu \nu}}{k^2-M_V^2} \nonumber\\
S(k) &=& \frac{i}{k^2-M_S^2} 
\en
for the vector and scalar propagators, respectively. The previously discussed 
arbitrary parameters $w_1$ and $w_2$ are now constrained to $w_1 = w_2 = 1/2$.
The vertex factors can be easily found by calculating the derivative of the 
Fourier transformed action of the corresponding diagram with respect to the fields
which are attached to the vertex
\eq
		i \Gamma_n(p_1, p_2, ..., p_n) = 
		\left.\frac{i \delta^n S_I[\Phi]}{\delta\tilde{\Phi}_1(p_1) 
		\delta\tilde{\Phi}_2(p_2) ... \delta\tilde{\Phi}_n(p_n)}
		\right|_{\tilde{\Phi}_i = 0}.
\en
The relevant vertices are denoted in Fig. \ref{diagramsVertices}. 
One finds for the vertex factors by dropping the usual factor 
$(2\pi)^4 \delta^4(\sum P_{\text{in}} - \sum P_{\text{out}})$:
 \begin{align*}
\text{(V1):}\,\, i\Gamma &= \frac{-i g_{Z_c}}{\sqrt{2}}m_{Z_c} 
\Phi(-z_1)g^{\alpha \mu} \\
	\text{(V2):}\,\, i\Gamma &= \frac{i g_{Z_c}}{2\sqrt{2}}m_{Z_c} 
e\,(p_1+p_2+q/2)^\rho g^{\alpha \mu} \\
&\,\times\int\limits_{0}^{1} dt \Phi^\prime(- z_2 t - z_1 (1-t)) \\
	\text{(V3):}\,\, i\Gamma 
&= i e \left[g^{\mu \nu}(p_1+p_2)^\rho-g^{\mu \rho}p_1^\nu-g^{\nu \rho}p_2^\mu\right] \\
					&\approx ie\, (p_1 + p_2)^\rho g^{\mu \nu} \\
	\text{(V4):}\,\, i\Gamma &= -  ie\, (p_1 + p_2)^\rho \\
	\text{(V5):}\,\, i\Gamma 
&= - i g_{D D^\ast J/\psi \pi}\, \sqrt{2}\, (p_2\cdot p_1 g^{\mu \beta}-p_2^\mu p_1^\beta) \\
	\text{(V6):}\,\, i\Gamma 
&=  i g_{D D^\ast J/\psi \pi}\, e \sqrt{2}\, (g^{\mu \beta}p_2^\rho-p_2^\mu g^{\beta \rho}) 
 \end{align*}
where $z_1 = (p_1/2+p_2/2)^2$\,, 
$z_2 = (q/2+p_1/2+p_2/2)^2$. 

\section{Gauge invariance} 
\label{gaugeinv}
In this Appendix we demonstrate that gauge invariance 
is fulfilled for the transition amplitude of the three-body decay $\zc\to \jp\pi^+\gam$. 
The $\bar{D}^{0*} D^+$-meson loop integrals corresponding to the diagrams 
in Fig. \ref{diagramsZcJpsPiGamDsD1} are given by 
(we drop the general constant 
$c = - \frac{i}{16 \pi^2} g_{Z_c}\, m_{Z_c}\, g_{D D^\ast J/\psi \pi}\, e$)
\begin{widetext}
\begin{align*}
	i I_{1a} &= -\left( g^{\alpha \beta} p_2^{\rho} - 
p_2^{\alpha} g^{\beta \rho}\right ) \int \frac{d^4k}{\pi^2 i}
\Phi(-k^2)\tilde{S}(k+p/2)\tilde{D}(k-p/2) \\
	i I_{2a} &= -\left(p_2 \cdot p_1 g^{\alpha \beta} - 
p_2^{\alpha} p_1^{\beta}\right ) \int \frac{d^4k}{\pi^2 i}
\Phi(-k^2)\tilde{S}(k+p/2)\tilde{S}(k+p/2-q)\tilde{D}(k-p/2)
						 (2k+p-q)^{\rho} \\
	i I_{3a} &= \left(p_2 \cdot p_1 g^{\alpha \beta} - 
p_2^{\alpha} p_1^{\beta}\right ) \int_{0}^{1} dt 
\int \frac{d^4k}{\pi^2 i} (k+3/4q)^{\rho}
\Phi^\prime(-(k+q)^2 t - (k+q/2)^2(1-t)) \tilde{D}(k-p/2+q)\tilde{S}(k+p/2) \\
	i I_{4a} &= -\left((p-p_2)\cdot p_2 g^{\alpha \beta} - 
p_2^{\alpha} (p-p_2)^{\beta}\right) (2p_1+q)^{\rho}  
\int \frac{d^4k}{\pi^2 i}\Phi(-k^2)
		 \tilde{S}(k+p/2)\tilde{D}(k-p/2)\tilde{S}_\pi(p-p_2) \\
	i I_{5a} &= -\left(p_2 \cdot p_1 g^{\alpha \beta} - 
p_2^{\alpha} p_1^{\beta}\right ) (2p-q)^{\rho} 
\int \frac{d^4k}{\pi^2 i}\Phi(-k^2) \tilde{S}(k+p/2-q/2)
		\tilde{D}(k-p/2+q/2)\tilde{D}_{Z_c}(p-q)
\end{align*}
\end{widetext}
where $k$ is the loop momentum and
\begin{align*}
	\tilde{S}_{\pi}(k) &= \frac{1}{m_\pi^2-k^2}, \, \, 
&\tilde{D}_{Z_c}(k) = \frac{1}{m_{Z_c}^2-k^2}, \\
	\tilde{S}(k) &= \frac{1}{m_D^2-k^2}, \, \, 
&\tilde{D}(k) = \frac{1}{m_{D^\ast}^2-k^2},
\end{align*}
are the propagators of the scalar and vector fields, respectively.
 As was already mentioned, the transverse parts of the vector 
propagators are neglected.

To test for gauge invariance every loop diagram is contracted with 
the photon momentum $q$. In the following it will be helpful 
to establish the following relations by taking advantage of momentum 
conservation $p = p_1 + p_2 + q$:
\eq
	q \cdot (2k +p-q) 
&=& (k+p/2)^2-(k+p/2-q)^2 \\
	&=& \tilde{S}^{-1}(k+p/2-q) \nonumber\\&\,&- \tilde{S}^{-1}(k+p/2) \label{help1}\\
	q \cdot (2p_1+q) 
	&=& -\tilde{S}_\pi^{-1}(p-p_2)\label{help2}\\
	q \cdot (2p-q) &=& \tilde{D}_{Z_c}^{-1}(p_1+p_2).
	\label{help3}
\en
Multiplying diagram $i I_{1a}$ with $q$ we get
\eq\label{qI1}
	q i I_{1a} = -\left( g^{\alpha \beta} q\cdot p_2 - 
p_2^{\alpha} q^{\beta}\right ) \nonumber\\ 
\int \frac{d^4k}{\pi^2 i}\Phi(-k^2)\tilde{S}(k+p/2)\tilde{D}(k-p/2).
\en 
For the contraction of $i I_{2a}$ with $q$ and using (\ref{help1}) we obtain
\eq\label{qI2}
	q i I_{2a} =  -\left(p_2 \cdot p_1 g^{\alpha \beta} - 
p_2^{\alpha} p_1^{\beta}\right )
\int \frac{d^4k}{\pi^2 i}\Phi(-k^2)\nonumber\\
\tilde{D}(k-p/2) \left[ \tilde{S}(k+p/2) - \tilde{S}(k+p/2-q) \right].
\en 
Diagram $i I_{3a}$ multiplied with $q$ reads as
\eq\label{qI3}
	q i I_{3a} &=& \left(p_2 \cdot p_1 g^{\alpha \beta} - 
p_2^{\alpha} p_1^{\beta}\right ) \int \frac{d^4k}{\pi^2 i} \Phi(-k^2)\nonumber\\
&\times&\left[\tilde{D}(k-p/2+q/2)\tilde{S}(k+p/2-q/2)\right.\nonumber\\ 
&-&\left.\tilde{D}(k-p/2)\tilde{S}(k+p/2-q)\right]
\en
where we have used 
\eq
\int_{0}^{1}dt\, q \cdot (k+3/4 q) \Phi^\prime(-(k+q)^2 t - 
(k+q/2)^2(1-t)) \nonumber\\= -\Phi(-(k+q)^2)+\Phi(-(k+q/2)^2) \,. 
\hspace{10mm}\nonumber
\en
Furthermore we have shifted the first term of the integral containing the 
latter expression by $k \rightarrow k - q$ and the second term by 
$k \rightarrow k-q/2$.
When multiplying $i I_{4a}$ with $q$ we obtain with the help of (\ref{help2}) 
and using momentum conservation
\eq\label{qI4}
	q i I_{4a} =	\left((p_1+q)\cdot p_2 g^{\alpha \beta} - 
p_2^{\alpha} (p_1+q)^{\beta}\right) \nonumber\\ \int \frac{d^4k}{\pi^2 i}\Phi(-k^2)
							 \tilde{S}(k+p/2)\tilde{D}(k-p/2).
\en
\newline
We get the last expression with (\ref{help3}) and by multiplying  
$i I_{5a}$ with $q$
\eq\label{qI5}
	q i I_{5a} =  -\left(p_2 \cdot p_1 g^{\alpha \beta} - 
p_2^{\alpha} p_1^{\beta}\right ) \int \frac{d^4k}{\pi^2 i}\Phi(-k^2)\nonumber\\ 
\tilde{S}(k+p/2-q/2) \tilde{D}(k-p/2+q/2).
\en
Now it is easy to show that the expressions (\ref{qI1}) to (\ref{qI5}) cancel
\eq
	q \left(i I_{1a}+i I_{2a}+i I_{3a}+i I_{4a}+i I_{5a}\right) = 0.
\en
Thus gauge invariance for 
the $\bar{D}^{0*} D^+$-meson loop integrals of Fig.~\ref{diagramsZcJpsPiGamDsD1} 
is shown. The proof of gauge invariance
for the $\bar{D}^{0} D^{*+}$ loop diagrams of Fig.~\ref{diagramsZcJpsPiGamDsD2} proceeds
exactly the same way as 
for the previous case of the $\bar{D}^{0*} D^+$-meson loop integrals by inserting the 
replacements $\tilde{S}(k) \leftrightarrow \tilde{D}(k)$.
\newline
\newline
\newline
\section{Structure integrals} 
\label{structint}
In this Appendix the structure integrals $F_1$, $F_{235} \equiv F_{2} + F_3 + F_5$ 
and $F_4$ are explicitly listed.\\
We define $c \equiv i g_{Z_c}\, m_{Z_c}\, g_{D D^\ast J/\psi \pi}\, e/(16 \pi^2)$, $s \equiv \Lambda^{-2}$ 
and $Q \equiv q$ for the three-body decay $\zc\to \jp\pi^+\gam$ and 
$Q \equiv k_3 + k_4$ for the four-body decay $\zc\to \jp\pi^+l^+l^-$.
\begin{widetext}
\begin{center}
\begin{minipage}[t]{1\textwidth}
	\begingroup
	\parfillskip=0pt
	\begin{minipage}[t]{0.35\textwidth}
		\eq
			i F_1 = 2 c \int_{0}^{\infty} dx dy 
        \frac{e^{-r_1^2/a_1 - M_1^2}}{a_1^2} \nonumber
		\en
	\end{minipage}%
	\hfill
	\begin{minipage}[t]{0.65\textwidth}
		\eq
			r_{1} &=& p(s/2 + y) \nonumber \\
			a_{1} &=& s + x + y  \nonumber\\
			M_1^2 &=& -m_{Z_c}^2 (s/4+y)+m_{D}^2y+m_{D^\ast}^2 x \nonumber
		\en
	\end{minipage}%
	\par\endgroup
\end{minipage}
\begin{minipage}[t]{1\textwidth}
	\begingroup
	\parfillskip=0pt
	\begin{minipage}[t]{0.35\textwidth}
		\eq
			i F_{2} = c \int_{0}^{\infty} dx dy dz 
\frac{s+2z}{a_{2}^3} e^{-r_{2}^2/a_{2}} \nonumber \\
			\times \left(e^{-M_{2a}^2}+e^{-M_{2b}^2}\right) \nonumber
		\en
	\end{minipage}%
	\hfill
	\begin{minipage}[t]{0.65\textwidth}
		\eq
			r_{2} &=& p(s/2 + z) + Q y\nonumber \\
			a_{2} &=& s + x + y + z \nonumber\\
			M_{2a}^2 &=& -m_{Z_c}^2 (s/4+z)+m_{D}^2(x+y)\nonumber\\
&\,& +m_{D^\ast}^2 z - Q^2 \, y\nonumber\\
			M_{2b}^2 &=& -m_{Z_c}^2 (s/4+z)+m_{D^\ast}^2(x+y)\nonumber\\
&\,& +m_{D}^2 z - Q^2 \, y\nonumber
		\en
	\end{minipage}%
	\par\endgroup
\end{minipage}
\begin{minipage}[t]{1\textwidth}
	\begingroup
	\parfillskip=0pt
	\begin{minipage}[t]{0.35\textwidth}
		\eq
			i F_{3} = -c 
\int_{0}^{\infty} dx dy \int_{0}^{1}dt \frac{s (x-y)}{2 a_{3}^3} e^{-r_{3}^2/a_{3}} \nonumber \\
			\times \left(e^{-M_{3a}^2}+e^{-M_{3b}^2}\right) \nonumber
		\en
	\end{minipage}%
	\hfill
	\begin{minipage}[t]{0.65\textwidth}
		\eq
			r_{3} &=& \frac{1}{4} Q \left( s (2 t -1) - 3x + y \right)  + \frac{1}{2} p(x- y) \nonumber \\
			a_{3} &=& s + x + y \nonumber\\
			M_{3a}^2 &=& (4 p \cdot Q (3x+y)-Q^2(s+9x+y)+\nonumber\\
							&\,& 4(4m_{D}^2 x+4 m_{D^\ast}^2 y-p^2(x+y)))/16 \nonumber\\
			M_{3b}^2 &=& (4 p \cdot Q (3x+y)-Q^2(s+9x+y)+\nonumber\\
							   		   &\,&4(4m_{D^\ast}^2 x+4 m_{D}^2 y-p^2(x+y)))/16 \nonumber
		\en
	\end{minipage}%
	\par\endgroup
\end{minipage}
\begin{minipage}[t]{1\textwidth}
	\begingroup
	\parfillskip=0pt
	\begin{minipage}[t]{0.35\textwidth}
		\eq
			i F_{4} = 4 c \int_{0}^{\infty} 
dx dy \frac{e^{-r_{4}^2/a_{4} - M_4^2}}{a_{4}^2}  \nonumber \\
			\times \frac{1}{m_\pi^2-(p_1 + Q)^2} \nonumber
		\en
	\end{minipage}%
	\hfill
	\begin{minipage}[t]{0.65\textwidth}
		\eq
			r_{4} &=& p (s/2 + y)\nonumber \\
			a_{4} &=& s + x + y \nonumber\\
			M_4^2 &=& -m_{Z_c}^2 (s/4+y)+m_{D}^2 y+m_{D^\ast}^2 x \nonumber
		\en
	\end{minipage}%
	\par\endgroup
\end{minipage}
\begin{minipage}[t]{1\textwidth}
	\begingroup
	\parfillskip=0pt
	\begin{minipage}[t]{0.35\textwidth}
		\eq
			i F_{5} = 4 c \int_{0}^{\infty} 
dx dy \frac{e^{-r_{5}^2/a_{5} - M_5^2} }{a_{5}^2}  \nonumber\\
			\times \frac{1}{m_Z^2-(p_1 + p_2)^2} \nonumber
		\en
	\end{minipage}%
	\hfill
	\begin{minipage}[t]{0.65\textwidth}
		\eq
			r_{5} &=& p (s/2 + x) + Q (s/2 + y)\nonumber \\
			a_{5} &=& s + x + y \nonumber\\
			M_5^2 &=& -m_{Z_c}^2 (s/4+x
)+m_{D^\ast}^2 x+m_{D}^2 y \nonumber\\ &\,& - p \cdot Q \, s/2 - Q^2\,(s/4+y) \nonumber
		\en
	\end{minipage}%
	\par\endgroup
\end{minipage}
\end{center}
\newpage
\section{Numerical values} 
\label{AppDecay}
We summarize the numerical results for 
the coupling constant $g_{Z_c}$ and 
the decay rates $\Gamma_3$, $\Gamma_4$ of the three- and four-body transitions.

\begin{table}[ht]
\def\arraystretch{.95}
	\setlength{\tabcolsep}{3mm}
	\caption{Numerical values for the dimensionless phenomenological 
	coupling constant $g_{Z_c}$  as a function of $\epsilon$ (column 1) and $\Lambda$ (row 1).}
	\begin{tabular}{||c|c|c|c|c|c|c|c||}
	\hline\hline
	$\epsilon\backslash\Lambda$ [MeV] & 500 & 550 & 600 & 650 & 700 & 750 & 800\\
	\hline\hline
    0.5 & 6.23 & 5.95 & 5.71 & 5.52 & 5.35 & 5.21 & 5.09 \\	
    1.0 & 6.36 & 6.07 & 5.83 & 5.62 & 5.45 & 5.31 & 5.18 \\	
    1.5 & 6.49 & 6.19 & 5.94 & 5.73 & 5.55 & 5.40 & 5.27 \\	
    2.0 & 6.62 & 6.31 & 6.05 & 5.83 & 5.65 & 5.49 & 5.36 \\	
    2.5 & 6.75 & 6.43 & 6.16 & 5.93 & 5.74 & 5.58 & 5.44 \\	
    3.0 & 6.88 & 6.54 & 6.27 & 6.03 & 5.84 & 5.67 & 5.53 \\	
    3.5 & 7.01 & 6.66 & 6.37 & 6.13 & 5.93 & 5.76 & 5.61 \\	
    4.0 & 7.13 & 6.77 & 6.48 & 6.23 & 6.03 & 5.85 & 5.70 \\	
    4.5 & 7.26 & 6.89 & 6.58 & 6.33 & 6.12 & 5.94 & 5.78 \\	
    5.0 & 7.38 & 7.00 & 6.69 & 6.43 & 6.21 & 6.02 & 5.86 \\
	\hline 
	\hline
	\end{tabular}
	\label{tabMassOp}
\def\arraystretch{.95}
	\setlength{\tabcolsep}{3mm}
	\caption{Numerical values for the decay rate $\Gamma_3$ in 
	$\unit{keV}$ for the transition $\zc\to \jp\pi^+\gamma$ for 
	$\delta m_\gamma = 150\unit{MeV}$ as a function of $\epsilon$ (column 1) and $\Lambda$ (row 1).}
	\begin{tabular}{||c|c|c|c|c|c|c|c||}
	\hline\hline
	$\epsilon\backslash\Lambda$ [MeV] & 500 & 550 & 600 & 650 & 700 & 750 & 800\\
	\hline\hline
    0.5 & 61.1 & 67.4 & 73.9 & 80.5 & 87.5 & 94.7 & 101.9 \\	
    1.0 & 58.1 & 64.2 & 70.8 & 77.5 & 84.5 & 91.7 &  99.2 \\	
    1.5 & 56.1 & 62.5 & 69.0 & 75.7 & 82.9 & 90.2 &  97.7 \\	
    2.0 & 54.9 & 61.3 & 67.9 & 74.8 & 81.8 & 89.3 &  96.8 \\	
    2.5 & 54.0 & 60.4 & 67.1 & 74.1 & 81.2 & 88.6 &  96.3 \\	
    3.0 & 53.3 & 59.8 & 66.5 & 73.6 & 80.8 & 88.3 &  96.1 \\	
    3.5 & 52.8 & 59.3 & 66.2 & 73.3 & 80.6 & 88.3 &  96.0 \\	
    4.0 & 52.4 & 59.1 & 65.9 & 73.1 & 80.6 & 88.1 &  96.1 \\	
    4.5 & 52.1 & 58.8 & 65.8 & 73.0 & 80.5 & 88.3 &  96.3 \\	
    5.0 & 51.9 & 58.7 & 65.7 & 73.0 & 80.6 & 88.4 &  96.5 \\
	\hline 
	\hline
	\end{tabular}
	\label{tabDecay1}
\def\arraystretch{.95}
	\setlength{\tabcolsep}{3mm}
	\caption{Numerical values for the decay rate $\Gamma_4$ 
	in $\unit{keV}$ for the transition $\zc\to \jp\pi^+e^+e^-$ 
	as a function of $\epsilon$ (column 1) and $\Lambda$ (row 1).}
	\begin{tabular}{||c|c|c|c|c|c|c|c||}
	\hline\hline
	$\epsilon\backslash\Lambda$ [MeV] & 500 & 550 & 600 & 650 & 700 & 750 & 800\\
	\hline\hline
	0.5 & 5.770 & 6.440 & 7.070 & 7.239 & 7.620 & 8.575 & 9.230\\
	1.0 & 4.802 & 5.118 & 5.725 & 6.079 & 6.664 & 7.149 & 7.597\\
	1.5 & 4.308 & 4.704 & 5.208 & 5.678 & 6.332 & 6.853 & 7.207\\
	2.0 & 4.123 & 4.531 & 4.970 & 5.484 & 5.995 & 6.536 & 6.999\\
	2.5 & 3.933 & 4.418 & 4.911 & 5.303 & 5.843 & 6.331 & 6.931\\
	3.0 & 3.826 & 4.308 & 4.782 & 5.321 & 5.859 & 6.293 & 6.885\\
	3.5 & 3.778 & 4.253 & 4.719 & 5.202 & 5.774 & 6.279 & 6.824\\
	4.0 & 3.726 & 4.174 & 4.673 & 5.174 & 5.704 & 6.246 & 6.812\\
	4.5 & 3.690 & 4.150 & 4.636 & 5.142 & 5.703 & 6.265 & 6.804\\
	5.0 & 3.661 & 4.141 & 4.623 & 5.151 & 5.699 & 6.278 & 6.854\\
	\hline 
	\hline
	\end{tabular}
	\label{tabDecay2}
\def\arraystretch{.95}
	\setlength{\tabcolsep}{3mm}
	\caption{Numerical values for the decay rate $\Gamma_4$ 
	in $\unit{eV}$ for the transition $\zc\to \jp\pi^+\mu^+\mu^-$ 
	as a function of $\epsilon$ (column 1) and $\Lambda$ (row 1).}
	\begin{tabular}{||c|c|c|c|c|c|c|c||}
	\hline\hline
	$\epsilon\backslash\Lambda$ [MeV] & 500 & 550 & 600 & 650 & 700 & 750 & 800\\
	\hline\hline
	0.5 & 9.593 & 10.559 & 11.574 & 12.621 & 13.711 & 14.826 & 15.986\\
	1.0 & 9.080 & 10.064 & 11.079 & 12.130 & 13.216 & 14.352 & 15.515\\
	1.5 & 8.779 & 9.7613 & 10.789 & 11.848 & 12.951 & 14.082 & 15.252\\
	2.0 & 8.572 & 9.5626 & 10.587 & 11.664 & 12.775 & 13.922 & 15.105\\
	2.5 & 8.415 & 9.4161 & 10.461 & 11.541 & 12.656 & 13.810 & 15.006\\
	3.0 & 8.297 & 9.3083 & 10.360 & 11.453 & 12.584 & 13.750 & 14.955\\
	3.5 & 8.206 & 9.2270 & 10.294 & 11.391 & 12.528 & 13.708 & 14.918\\
	4.0 & 8.136 & 9.1672 & 10.237 & 11.349 & 12.493 & 13.683 & 14.906\\
	4.5 & 8.084 & 9.1159 & 10.196 & 11.316 & 12.478 & 13.673 & 14.906\\
	5.0 & 8.035 & 9.0827 & 10.165 & 11.300 & 12.466 & 13.675 & 14.920\\
	\hline 
	\hline
	\end{tabular}
	\label{tabDecay3}
\end{table}
\end{widetext}


\newpage 
\newpage
\begin{thebibliography}{99}

\bibitem{Ablikim:2013mio}
  M.~Ablikim {\it et al.}  [BESIII Collaboration],
  Phys.\ Rev.\ Lett. {\bf 110}, 252001 (2013) 
  [arXiv:1303.5949 [hep-ex]].

\bibitem{Liu:2013dau}
  Z.~Q.~Liu {\it et al.}  [Belle Collaboration],
  Phys.\ Rev.\ Lett. {\bf 110}, 252002 (2013) 
  [arXiv:1304.0121 [hep-ex]].

\bibitem{Xiao:2013iha}
  T.~Xiao, S.~Dobbs, A.~Tomaradze and K.~K.~Seth,
  Phys.\ Lett.\ B {\bf 727}, 366 (2013)
  arXiv:1304.3036 [hep-ex].

\bibitem{Faccini:2013lda} 
  L.~Maiani, V.~Riquer, R.~Faccini, F.~Piccinini, A.~Pilloni and A.~D.~Polosa,
  Phys.\ Rev.\ D {\bf 87}, 111102 (2013)
  [arXiv:1303.6857 [hep-ph]].

\bibitem{Dias:2013xfa} 
  J.~M.~Dias, F.~S.~Navarra, M.~Nielsen and C.~M.~Zanetti,
  Phys.\ Rev.\ D {\bf 88}, 016004 (2013)
  [arXiv:1304.6433 [hep-ph]].

\bibitem{Braaten:2013boa} 
  E.~Braaten,
  Phys.\ Rev.\ Lett.\  {\bf 111}, 162003 (2013)
  [arXiv:1305.6905 [hep-ph]].

\bibitem{Wang:2013cya} 
  Q.~Wang, C.~Hanhart and Q.~Zhao,
  Phys.\ Rev.\ Lett.\  {\bf 111}, 132003 (2013)
  [arXiv:1303.6355 [hep-ph]].

\bibitem{Cui:2013yva} 
  C.~Y.~Cui, Y.~L.~Liu, W.~B.~Chen and M.~Q.~Huang,
  J.\ Phys.\ G {\bf 41}, 075003 (2014)
  [arXiv:1304.1850 [hep-ph]].

\bibitem{Zhang:2013aoa} 
  J.~R.~Zhang,
  Phys.\ Rev.\ D {\bf 87}, 116004 (2013)
  [arXiv:1304.5748 [hep-ph]].

\bibitem{Ke:2013gia} 
  H.~W.~Ke, Z.~T.~Wei and X.~Q.~Li,
  Eur.\ Phys.\ J.\ C {\bf 73}, 2561 (2013)
  [arXiv:1307.2414].

\bibitem{Dong:2013iqa}
  Y.~Dong, A.~Faessler, T.~Gutsche and V.~E.~Lyubovitskij,
  Phys.\ Rev.\ D {\bf 88}, 014030 (2013) 
  [arXiv:1306.0824 [hep-ph]].

\bibitem{Dong:2012hc} 
  Y.~Dong, A.~Faessler, T.~Gutsche and V.~E.~Lyubovitskij,
  J.\ Phys.\ G {\bf 40}, 015002 (2013)
  [arXiv:1203.1894 [hep-ph]].


\bibitem{Dong:2013kta} 
  Y.~Dong, A.~Faessler, T.~Gutsche and V.~E.~Lyubovitskij,
  Phys.\ Rev.\ D {\bf 89}, 034018 (2014)
  [arXiv:1310.4373 [hep-ph]].

\bibitem{Faessler:2007gv} 
  A.~Faessler, T.~Gutsche, V.~E.~Lyubovitskij and Y.~-L.~Ma,
  Phys.\ Rev.\ D {\bf 76}, 014005 (2007)
  [arXiv:0705.0254 [hep-ph]];  
  T.~Branz, T.~Gutsche and V.~E.~Lyubovitskij,
  Phys.\ Rev.\ D {\bf 80}, 054019 (2009)
  [arXiv:0903.5424 [hep-ph]]; 
  Phys.\ Rev.\ D {\bf 82}, 054025 (2010)
  [arXiv:1005.3168 [hep-ph]]; 
  Phys.\ Rev.\ D {\bf 78}, 114004 (2008)
  [arXiv:0808.0705 [hep-ph]]; 
  Y.~B.~Dong, A.~Faessler, T.~Gutsche and V.~E.~Lyubovitskij,
  Phys.\ Rev.\ D {\bf 77}, 094013 (2008)
  [arXiv:0802.3610 [hep-ph]];  
  Y.~Dong, A.~Faessler, T.~Gutsche and V.~E.~Lyubovitskij,
  arXiv:1404.6161 [hep-ph].

\bibitem{Dong:2009uf}
 Y.~Dong, A.~Faessler, T.~Gutsche and V.~E.~Lyubovitskij,
  J.\ Phys.\ G {\bf 38}, 015001 (2011)
  [arXiv:0909.0380 [hep-ph]]; 

\bibitem{Weinberg:1962hj}
  S.~Weinberg,
  Phys.\ Rev.\  {\bf 130}, 776 (1963).

\bibitem{Efimov:1993ei}
  G.~V.~Efimov and M.~A.~Ivanov,
  {\it The Quark Confinement Model of Hadrons},
  (IOP Publishing, Bristol $\&$ Philadelphia, 1993)

\bibitem{Branz:2009cd} 
  T.~Branz, A.~Faessler, T.~Gutsche, M.~A.~Ivanov, J.~G.~Korner and 
  V.~E.~Lyubovitskij,
  Phys.\ Rev.\ D {\bf 81}, 034010 (2010)
  [arXiv:0912.3710 [hep-ph]].  

\bibitem{Dong:2009tg} 
  Y.~Dong, A.~Faessler, T.~Gutsche and V.~E.~Lyubovitskij,
  Phys.\ Rev.\ D {\bf 81}, 014006 (2010)
  [arXiv:0910.1204 [hep-ph]]. 

\bibitem{Anikin:1995cf}
  I.~V.~Anikin, M.~A.~Ivanov, N.~B.~Kulimanova and V.~E.~Lyubovitskij,
  Z.\ Phys.\ C {\bf 65}, 681 (1995).  

\bibitem{Ivanov:1996pz}
  M.~A.~Ivanov, M.~P.~Locher and V.~E.~Lyubovitskij,
  Few-Body Syst.\  {\bf 21}, 131 (1996) 
  [hep-ph/9602372].
 
\bibitem{Ivanov:1996fj}
  M.~A.~Ivanov, V.~E.~Lyubovitskij, J.~G.~Korner and P.~Kroll,
  Phys.\ Rev.\ D {\bf 56}, 348 (1997) 
  [hep-ph/9612463].
   
\bibitem{Faessler:2001mr}
  A.~Faessler, T.~Gutsche, M.~A.~Ivanov, J.~G.~Korner and V.~E.~Lyubovitskij,
  Phys.\ Lett.\ B {\bf 518}, 55 (2001)  
  [hep-ph/0107205].
      
\bibitem{Faessler:2006ft}
  A.~Faessler, T.~Gutsche, M.~A.~Ivanov, J.~G.~Korner, V.~E.~Lyubovitskij, 
  D.~Nicmorus and K.~Pumsa-ard,
  Phys.\ Rev.\ D {\bf 73}, 094013 (2006)   
  [hep-ph/0602193].
        
    \bibitem{Faessler:2006ky}
      A.~Faessler, T.~Gutsche, B.~R.~Holstein, V.~E.~Lyubovitskij, D.~Nicmorus 
      and K.~Pumsa-ard,
      Phys.\ Rev.\ D {\bf 74}, 074010 (2006)   
      [hep-ph/0608015].

\bibitem{Mandelstam:1962mi}
  S.~Mandelstam,
  Ann. Phys. (N.Y.)\  {\bf 19}, 1 (1962).

\bibitem{Terning:1991yt}
  J.~Terning,
  Phys.\ Rev.\ D {\bf 44}, 887 (1991).  
  
\bibitem{Faessler:2003yf}
  A.~Faessler, T.~Gutsche, M.~A.~Ivanov, V.~E.~Lyubovitskij and P.~Wang,
  Phys.\ Rev.\ D {\bf 68}, 014011 (2003)  
  [hep-ph/0304031].

\bibitem{Faessler:2007us}
  A.~Faessler, T.~Gutsche, V.~E.~Lyubovitskij and Y.~-L.~Ma,
  Phys.\ Rev.\ D {\bf 76}, 114008 (2007) 
  [arXiv:0709.3946 [hep-ph]].

\bibitem{Faessler:2007cu}
  A.~Faessler, T.~Gutsche, S.~Kovalenko and V.~E.~Lyubovitskij,
  Phys.\ Rev.\ D {\bf 76}, 1 (2007) 
  [arXiv:0705.0892 [hep-ph]].

\bibitem{Byckling:1973}
  E.~Byckling~and~K.~Kajantie, 
 \textit{Particle Kinematics}, (Wiley, New York, 1973), p. 319. 

\bibitem{Cabibbo:1965zzb}
  N.~Cabibbo and A.~Maksymowicz,
  Phys.\ Rev.\  {\bf 137}, B438 (1965); 
   {\bf 168}, 1926 [E] 1968.

\bibitem{Bijnens:1992mk} 
  J.~Bijnens, G.~Ecker and J.~Gasser,
        in \textit{The DA$\Phi$NE Physics Handbook},
	edited\ by\ L.\ Maiani,\ G.\ Pancheri,\ and\ N.\ Paver\ 
        (Servizio Documentazione dei Laboratori Nazionali 
        di Frascati, Frascati, 1992), Vol. I. 

\bibitem{Daphne:1992}
	V.S.~Demidov~and~E.~Shabalin,
        in \textit{The DA$\Phi$NE Physics Handbook},
	edited\ by\ L.\ Maiani,\ G.\ Pancheri,\ and\ N.\ Paver\ 
        (Servizio Documentazione dei Laboratori Nazionali 
        di Frascati, Frascati, 1992), Vol. I

\bibitem{Bijnens:1993xi}
  J.~Bijnens,
  Int.\ J.\ Mod.\ Phys.\ A {\bf 8} 3045 (1993).


\bibitem{Knochlein:1996ah}
  G.~Knochlein, S.~Scherer and D.~Drechsel,
  Phys.\ Rev.\ D {\bf 53} (1996) 3634
  [hep-ph/9601252].

\bibitem{Dubnicka:2010kz} 
  S.~Dubnicka, A.~Z.~Dubnickova, M.~A.~Ivanov and J.~G.~Korner,
  Phys.\ Rev.\ D {\bf 81}, 114007 (2010)
  [arXiv:1004.1291 [hep-ph]]; 
  S.~Dubnicka, A.~Z.~Dubnickova, M.~A.~Ivanov, J.~G.~Korner,
  P.~Santorelli and G.~G.~Saidullaeva,
  Phys.\ Rev.\ D {\bf 84}, 014006 (2011) 
  [arXiv:1104.3974 [hep-ph]].


\end{thebibliography}
\end{document}